\documentclass[aps,twocolumn,showpacs,preprintnumbers,amsmath,amssymb,nofootinbib,superscriptaddress,showkeys]{revtex4-1}

\usepackage{epsfig}
\usepackage{graphicx}
\def\Xint#1{\mathchoice
{\XXint\displaystyle\textstyle{#1}} 
{\XXint\textstyle\scriptstyle{#1}} 
{\XXint\scriptstyle\scriptscriptstyle{#1}} 
{\XXint\scriptscriptstyle\scriptscriptstyle{#1}} 
\!\int}
\def\XXint#1#2#3{{\setbox0=\hbox{$#1{#2#3}{\int}$ }
\vcenter{\hbox{$#2#3$ }}\kern-.5\wd0}}

\def\dashint{\Xint-}

\def\slashchar#1{\setbox0=\hbox{$#1$}
   \dimen0=\wd0 \setbox1=\hbox{/} \dimen1=\wd1
   \ifdim\dimen0>\dimen1 \rlap{\hbox to \dimen0{\hfil/\hfil}} #1
   \else  \rlap{\hbox to \dimen1{\hfil$#1$\hfil}} / \fi}

\begin{document}

\title{Meson dominance of hadron form factors and large-$N_c$ phenomenology}\thanks{Supported by
MICINN of Spain (FPA2010-16802, FPA2010-16696, FIS2011-24149), Consolider-Ingenio 2010 Programme CPAN
(CSD2007-00042), Junta de Andaluc\'{\i}a
(FQM 101, FQM 437, FQM225 and  FQM022), by the Deutsche Forschungsgemeinschaft DFG through the Collaborative Research Center ``The Low-Energy Frontier of the Standard Model" (SFB 1044), the Polish Science and Higher Education, grant N~N202 263438, and by National Science
Centre, grant DEC-2011/01/B/ST2/03915.}
\author{Pere Masjuan} \email{masjuan@kph.uni-mainz.de}
\affiliation{Departamento de F\'{i}sica Te\'{o}rica y del Cosmos and CAFPE,
Universidad de Granada, E-18071 Granada, Spain} \affiliation{Institut f\"ur Kernphysik, Johannes Gutenberg-Universit\"at, D-55099 Mainz, Germany }

\author{Enrique Ruiz Arriola} \email{earriola@ugr.es}
\affiliation{Departamento de F\'isica At\'omica, Molecular y Nuclear
  and Instituto Carlos I de Fisica Te\'orica y Computacional,
  Universidad de Granada, E-18071 Granada, Spain}

\author{Wojciech Broniowski} \email{Wojciech.Broniowski@ifj.edu.pl}
\affiliation{The H. Niewodnicza\'nski Institute of Nuclear Physics,
  PL-31342~Krak\'ow, Poland} \affiliation{Institute of Physics, Jan
  Kochanowski University, PL-25406~Kielce, Poland}

\date{\today}

\begin{abstract} 
We discuss the pion and nucleon form factors and generalized form
factors within the large-$N_c$ approach in the space-like region.  We
estimate their theoretical uncertainties through the use of the {\em
  half-width rule}, amounting to taking the half-width of the
resonances as the deviation of their mass parameters.  The approach
embodies the meson dominance of form factors and the high-energy
constraints from perturbative QCD. We compare the results with the
available experimental data and lattice simulations. The
meson-dominance form factors are generally comparable to the available
experimental data within the half-width-rule uncertainties. Our errors
are comparable to the experimental uncertainties, but are smaller than
lattice errors.
\end{abstract}

\pacs{12.38.Lg, 11.30, 12.38.-t} 

\keywords{meson form factors, large $N_c$, resonances, meson dominance}

\maketitle

\section{Introduction}

Electroweak form factors provide valuable information on the internal
structure of the existing composite hadrons, particularly on their
Lorentz-invariant transverse densities~\cite{Miller:2010nz} (for
reviews of the nucleon form factors see, e.g.,
Ref.~\cite{Drechsel:2007sq} and references therein).  On the
fundamental level, the QCD counting rules provide the high-momentum
behavior of the form factors~\cite{Brodsky:1976rz}, which in
perturbative QCD (pQCD) acquire logarithmic
corrections~\cite{Lepage:1980fj}. In recent years much attention has also
been paid to the so-called generalized form factors, i.e., moments of
the generalized parton distributions (GPDs)~(for reviews see
e.g.~\cite{Radyushkin:2000uy,Goeke:2001tz,Ji:2004gf} and references
therein, which can be directly accessed on the Euclidean
lattices~\cite{Hagler:2009ni}, even when no physical experiments can.

Electromagnetic form factors are by far best understood, both
theoretically and experimentally. An early and phenomenologically very
successful approach to study these quantities was initiated by the
Vector Meson Dominance (VMD) model (for reviews see
\cite{sakurai1969currents,O'Connell:1995wf}). On a formal level, VMD
is implemented in two equivalent ways. At the field-theoretic level
one postulates the so-called current-field identities which state the
proportionality between fields of stable mesons and the related
conserved currents with identical quantum numbers. Alternately, within
the framework of dispersion relations, one may saturate the matrix
elements of the currents with delta-like spectral functions. However,
vector mesons are unstable resonances with a finite decay width, thus
corrections to the narrow resonance approximation are expected.

There arises a natural question on {\em what} numerical value for the
mass of the resonance one should
use~\cite{Masjuan:2008fv,Masjuan:2008xg,Masjuan:2012wy}. While the
rigorous definition of a resonance mass $m_R$ and width $\Gamma_R$
corresponds to a pole of an analytically continued amplitude in the
complex s-plane on the second Riemann sheet, $s=m_R^2 -i \Gamma_R
m_R$, complex energies cannot be measured experimentally.  Of course,
since a resonance has a real mass distribution which generally depends
on the process where the resonance is produced, a variety of
possibilities arise to extract the maximum of an integrated mean value,
which becomes identical and independent of the background in the narrow-resonance 
limit. Thus, a conservative estimate is made if the mass of
a resonance is determined with an accuracy of about its half
width. Following
Refs.~\cite{RuizArriola:2010fj,Arriola:2010aj,Arriola:2011en,Masjuan:2012gc,Masjuan:2012yw}
we suggest to use the {\em half-width rule}: take the intrinsic
uncertainty of a resonance mass as the range $m_R \pm \Gamma_R/2$ to
estimate the finite width corrections. The average width-to-mass ratio
listed in the PDG was found to be $\langle \Gamma_R/m_R \rangle =0.12(8)$, both for
mesons and baryons~\cite{Arriola:2011en,Masjuan:2012gc}.

From a fundamental point of view, the success of the VMD model
remained a mystery until it was shown how it arises within a well
defined approximation of QCD. Indeed, in the large-$N_c$ limit
resonances become narrow~\cite{'tHooft:1973jz,Witten:1979kh} and
hadronic form factors turn out to be meromorphic functions under the
assumption of confinement; although in principle they contain an
infinite set of resonances, they can generally be written as a sum of
pole
functions~\cite{Knecht:1997ts,Peris:1998nj,Pich:2002xy,Masjuan:2007ay,Masjuan:2008fr}.
This allows us to build up an effective theory at a purely hadronic
level, with no explicit reference to the underlying quark-gluon
dynamics: all the specific QCD information is contained in the chiral and
short-distance constraints.  More generally, meson dominance of any
vertex function with appropriate quantum numbers and the meromorphic
property extends also to the case of generalized form factors. From a
practical point of view, for the lowest-rank generalized form factors
the only difference from the standard form factors associated to
conserved currents is that while the momentum dependence remains scale
independent, the normalization undergoes the QCD
evolution~\cite{Broniowski:2009zh}. Thus, generalized form factors in
the space-like region provide nothing but meson masses estimated in
the large-$N_c$ limit.
 
It is quite remarkable that given the simplicity of the leading-$N_c$
contribution, where only tree diagrams are needed, the $1/N_c$
corrections turn out to be extremely complicated and despite
courageous
efforts~\cite{Rosell:2004mn,Rosell:2005ai,Rosell:2006dt,Pich:2010sm,Pich:2008jm}
they have not been worked out completely. In the present paper we
provide a simple way of estimating the size of the $1/N_c$ corrections
by implementing a rather obvious idea that the mass of a resonance is
determined with an accuracy of about its half
width~\cite{RuizArriola:2010fj,Arriola:2010aj,Arriola:2011en,Masjuan:2012gc,Masjuan:2012yw}.
While this provides a large-$N_c$ meaning to the half-width rule, it
also yields rather rewarding consequences; the predicted theoretical
uncertainties turn out to be comparable or larger than the
corresponding experimental results, but at the same time smaller than
current lattice estimates.  In all considered cases we find an overall
agreement between the theory and experiment.

The requirement of quark-hadron duality at large $N_c$ involves, as a
matter of principle, an infinite tower of hadronic
states~\cite{Golterman:2001nk,Beane:2001uj,Afonin:2004yb,RuizArriola:2006gq,Arriola:2006sv,Masjuan:2007ay,Masjuan:2008fr,Queralt:2010sv}. 
This becomes clear for two-point functions, where further asymptotic
constraints between meson, the decay amplitudes and the meson spectra are
derived~\footnote{In fact, the only possibility to avoid the dimension-2
  operators at large $Q^2$, not present in the conventional Operator
  Product Expansion, is by assuming an infinite set of
  states~\cite{RuizArriola:2006gq,Arriola:2006sv}.}.  A careful
scrutiny of the Particle Data Table confirms, with the help of the half-width rule~\cite{Masjuan:2012gc}, 
the radial and angular momentum
Regge pattern in the meson spectrum, proposed in
Ref.~\cite{Anisovich:2000kxa}. This fact provides a phenomenological basis for
large-$N_c$ Regge model calculations of form
factors~\cite{Dominguez:2001zu,RuizArriola:2006ii,RuizArriola:2008sq,RuizArriola:2010fj}.

Unlike for the two-point functions, the short-distance QCD constraints for the
form factors may be saturated with a finite number of states, provided a detailed
pQCD information (the occurrence of the logarithmic corrections) is given up. 
One may take advantage of this fact by using
a sufficiently large but finite number of meson states, such that the correct
asymptotics is reproduced up to (slowly varying) logarithms, which allows
to impose the appropriate normalization conditions~\cite{Pich:2002xy,Bijnens:2003rc,Masjuan:2007ay,Masjuan:2008fr,Masjuan:2012wy}. The
implementation of the pQCD logs from the mesonic side is not at all
trivial; a possible mechanism, involving infinitely many states, is suggested in
Ref.~\cite{RuizArriola:2008sq}, where it was also shown that the onset of
pQCD in the pion form factor might possibly occur at ``cosmologically''
large momenta.  We recall in this regard that
almost model-independent upper and lower bounds on the spacelike form
factor are established above $Q^2 \sim 7~{\rm GeV}^2$~\cite{Ananthanarayan:2012tn}. To be fair, it is
not completely clear whether at present the logarithmic pQCD corrections are
distinctly seen in the current experimental data.
We note that approaches with an infinite number of meson resonances are considered
along the holographic framework~\cite{Hong:2007dq,Harada:2010cn}.

Usually, nucleon form factors are conveniently parameterized as dipole
functions, which describe the data quite successfully in a given range
of momenta.  However, this can only correspond {\em exactly} to a
sum of two simple  degenerate poles with opposite residues.  Actually, we will
show that if the error bars on the monopole mass are taken into
account, one can make the dipole overlap with a product of monopoles
within the corresponding error bars provided with the half-width rule.

To summarize, our construction is based on the following assumptions:
\begin{itemize}
\item Hadronic form factors in the space like region are dominated by
  mesonic states with the relevant quantum numbers.
\item The high-energy behavior is given by pQCD, and the number of
  mesons is taken to be minimal to satisfy these conditions.
\item Errors in the meson-dominated form factors are estimated by
  means of the half-width rule, i.e.,  by treating resonance masses as
  random variables distributed with the dispersion given by the width.
\end{itemize}

The paper is organized as follows. In Section \ref{sec:mes-dom} we
review for completeness the basics of meson dominance in the narrow
width limit for the case of the nucleon. In Section
\ref{sec:fin-width} we digress on the role played by the finite-width
corrections, providing a large-$N_c$ justification for the intuitively
obvious half-width rule for the masses. In Sections \ref{sec:pion-FF}
and \ref{sec:nuclen-FF} we carry out the analysis for several pion and
nucleon form factors, respectively. Finally, in Section \ref{sec:concl}
we draw our main conclusions.

\section{Meson dominance of form factors}
\label{sec:mes-dom}

First, the implications of meson dominance on form factors will be illustrated
with the nucleon form factors as an example. Quite generally, the
nucleon form factors are defined as the matrix element of a given
current or composite interpolating field, $J(x)$,
\begin{eqnarray}
\langle N(p',s') | J (0) | N(p,s) \rangle = \bar u (p',s') \Gamma_J
(p'-p) u (p,s) , \nonumber \\
\end{eqnarray} 
where $u(p,s)$ and $u(p',s')$ are Dirac spinors corresponding to the
initial and final four-momentum and spin states, respectively. 
The quantity $\Gamma_J(p'-p)$ involves the Dirac matrices. Lorentz
indices are suppressed for clarity of notation. Meson dominance of
the form factor corresponds to parameterizing 
the current with a superposition of meson fields with the same quantum
numbers as the current, 
\begin{eqnarray}
J (x) = \sum_n f_n \Phi_n (x),
\end{eqnarray} 
which means that the meson may decay into the vacuum through the
current, 
\begin{eqnarray}
\langle 0 | J(0) | \Phi_n \rangle = f_n .  
\end{eqnarray} 
This also implies that the two-point correlator 
can be written as 
\begin{eqnarray}
\Pi_{JJ} (t ) = \sum_n \frac{f_n^2}{m_n^2-t} \, ,
\label{eq:tpf}
\end{eqnarray} 
where $m_n$ is the mass of the meson state $\Phi_n $.
%(*** I tried to fix the notation below, please, check carefully. ***)
On the other hand, the meson-nucleon-nucleon coupling $g_n$ is defined via
\begin{eqnarray}
 \langle N(p',s') | (\partial^2 +M_n^2) \Phi_n (0) | N(p,s) \rangle =  \bar u(p',s') g_n u(p,s) \nonumber \\
\end{eqnarray} 
($g_n$ in general involves a Dirac structure).
Then 
\begin{eqnarray}
&& \langle N(p',s') | J (0) | N(p,s) \rangle = \sum_n f_n \langle N(p',s') | \Phi_n (0) |
N(p,s) \rangle \nonumber \\  
&&  = \bar u(p') F(t) u(p) ,
\end{eqnarray} 
where the form factor is
\begin{eqnarray}
F(t) = \sum_n \frac{f_n g_n}{m_n^2-t}. 
\label{eq:FFMD}
\end{eqnarray} 
It satisfies, on very general field-theoretic grounds and up to suitable subtractions, a dispersion
relation 
\begin{eqnarray}
F(t) = {\rm c.t.}+\frac1\pi \int_{t_0}^\infty \frac{{\rm Im}
F(t')}{t'-t-i\epsilon} dt' ,
\end{eqnarray} 
where c.t. stands for counterterms and $t_0$ is the threshold. In the narrow-resonance approximation one has the spectral density 
\begin{eqnarray}
{\rm Im} F(t) = \pi\sum_n c_n m_n^2 \delta (t-m_n^2),
\end{eqnarray} 
yielding, up to subtractions, the sum of monopoles  
\begin{eqnarray}
F(t) = \sum_n c_n \frac{m_n^2}{m_n^2-t}, 
\label{eq:ff_monopoles}
\end{eqnarray} 
corresponding to Eq.~(\ref{eq:FFMD}) with $c_n m_n^2=f_n g_n$ which is 
constant, i.e., independent of $t$.

The asymptotic behavior of form factors determines the number of necessary 
subtractions. Thus, for a form factor falling off as $\sim t^{-N}$ we
have a set of conditions
\begin{eqnarray}
\sum_n c_n m_n^k =0  \qquad k=0,...N-1\, ,
\end{eqnarray} 
and thus the minimum number of meson states needed to satisfy these constrains is $N$, whence 
\begin{eqnarray}
F(t) = F(0) \prod_{n=1}^N \frac{m_n^2}{m_n^2-t}. \label{eq:su}
\end{eqnarray} 
This simple ansatz predicts already the couplings in
Eq.~(\ref{eq:FFMD}) ``for free''. One can, of course, add more resonances
by multiplying Eq.~(\ref{eq:su}) with a factor
\begin{eqnarray}
\frac{1- d_k t/m_k^2}{1-t/m_k^2},
\end{eqnarray} 
where the unknown coefficient $d_k$ may be determined if some
couplings $c_n$ are known from the experiment. 

Quite generally, on the basis of the large-$Q^2$ expansion, one
has~\cite{Lepage:1980fj} $ (-t)^{i+1} F_i (t) \sim \log
(-t/\Lambda)^{-\gamma}$, with the anomalous dimension $\gamma \sim 2$
and weakly depending on the number of flavors.  Fits of the 
form factors hardly see
any impact of these pQCD logs at the currently available
momenta~\cite{Belushkin:2006qa}. 

The corresponding radii are given by the expansion 
\begin{eqnarray}
\frac{F(t)}{F(0)}= 1 + \frac{t}{6} \langle r^2 \rangle + \dots 
\end{eqnarray}
We recall that the radii are quite sensitive to chiral ($1/N_c$-%
suppressed) corrections and, actually, in some channels they diverge
for $m_\pi \to 0$.

Both two- and three-point correlators, Eqs.~(\ref{eq:tpf}) and
(\ref{eq:ff_monopoles}) respectively, require in principle an infinite
number of mesons.  Note, however, that the sign of the residues $f_n
g_n$ appearing in the form factors is arbitrary, while the sign
appearing in the two-point correlator is positive. This possibly
provides a quite different mechanism for cancellations, and hence for
the form how the short distance constraints are fulfilled. In short,
the two point functions {\it need} infinitely many mesonic states to
comply to pQCD, whereas the three-point functions, such as the form
factors, can be saturated with a finite number of meson states.

It is useful to notice that we may also  deduce the 
component of the $NN$ potential due to the exchange of the meson
states $\Phi_n$, 
\begin{eqnarray}
&& \langle N (p_1',s_1') N (p_2',s_2') | V|
N (p_1,s_1) N (p_2,s_2) \rangle =  \\ 
&& \sum_n  \bar u (p_1',s_1') g_n u (p_1,s_1)  
\bar u (p_2',s_2') g_n u (p_2,s_2) \frac{1}{m_n^2-t}. \nonumber  
\end{eqnarray} 
Via crossing, the $\bar NN $ scattering amplitude can be
obtained as well.

\section{Finite width corrections}
\label{sec:fin-width}

A question of fundamental and practical importance is what mass value
should one use for the meson states in the VMD expression for the form
factors~\cite{Masjuan:2008fv,Masjuan:2012gc}. Naively, one might take
the ``experimental'' value\footnote{This value also depends on the
  experimental process and may differ within the half-width
  rule.}. However, the extended VMD formula for the form factor,
Eq.~(\ref{eq:ff_monopoles}), corresponds to the large-$N_c$ limit. As
such, it is subject to the $1/N_c$ corrections which generate a
corresponding mass shift. The form of these corrections can in
principle be evaluated by computing meson loops within the Resonance
Chiral Perturbation
Theory~\cite{Rosell:2004mn,Rosell:2005ai,Rosell:2006dt,Pich:2010sm,Pich:2008jm,Portoles:2010yt}. The
general structure of the correction for the form factor corresponds to
the replacements
\begin{eqnarray}
\frac{g_n f_n }{m_n^2-t} \to \frac{G_n (t) f_n }{m_n^2-t - \Sigma (t)},
\end{eqnarray}
where $\Sigma (t)$ is the self-energy. However, the question remains what the size of
these corrections is numerically. Strictly speaking, such a question can
only be answered by a lattice calculation at different values of $N_c$
(see e.g. Ref.~\cite{Teper:2009uf} and references therein). Unfortunately, 
as we argue below, within a purely hadronic resonance theory we can only
make an educated guess, since there are undetermined counterterms
encoding the effects of the high-energy states not considered explicitly. For
instance, for the case of the $\rho$-meson we may take into account
the decay into $2\pi$, which is a real process, but also the virtual
$\bar K K$ excitation, etc. Our lack of an explicit knowledge on {\em all}
excitations makes it difficult, if not impossible, to predict the mass
shift reliably.

\subsection{Mass shift and the width}

To elaborate on the mass-shift effect in a greater detail, let us consider the
two-point function yielding the mesonic resonance propagator,
\begin{eqnarray}
D(s)= \frac{1}{s-m_0^2-\Sigma(s)}.
\end{eqnarray}
The mass parameter $m_0$ is the tree level resonance mass, which is
${\cal O} (N_c^0)$, whereas the self energy, coming from meson loops, is suppressed,  $\Sigma(s)={\cal O}
(N_c^{-1})$.  

Let us consider, for instance, the self-energy correction of the scalar
or vector mesons due to pion loops. Analyticity implies that the self-energy
satisfies a dispersion relation\footnote{We disregard spin
  complications, see, e.g.,~\cite{Kampf:2009jh} for details.}
\begin{eqnarray}
\Sigma(s) = {\rm c.t.} + \frac1{\pi} \int_{4 m_\pi^2}^\infty ds' \frac{{\rm Im} \Sigma(s'+i0^+)}{s'-s},
\label{eq:self-disp}
\end{eqnarray}
where c.t. means suitable subtractions. The pole position, $s=s_R = m_R^2-i m_R \Gamma_R$, is given by
\begin{eqnarray}
s_R - m_0^2 - \Sigma (s_R)=0  .
\end{eqnarray}
This is a  complicated self-consistent equation, but within the $1/N_c$ expansion 
it can be solved perturbatively, yielding
\begin{eqnarray}
s_R = m_0^2 + 2 m_0 \Delta m_R - i \Gamma_R m_0 + {\cal O} (N_c^{-2}),
\end{eqnarray}
where
\begin{eqnarray}
\Delta m_R &=& \frac1{2m_0} {\rm Re} \Sigma(m_0^2) ,  \\ 
%= \dashint \\ 
\Gamma_R &=& -\frac1{m_0} {\rm Im} \Sigma(m_0^2) .
\end{eqnarray}
Note that the imaginary part is proportional to the corresponding
decay width,

In general, there appears a threshold momentum dependence for the decay
amplitude which is proportional to the phase space. The form reflects the spin
of the resonance, such as 
\begin{eqnarray}
\Gamma(s)= \Gamma_R \left[\frac{\rho(s)}{\rho(m_R^2)}\right]^{2J+1},
\end{eqnarray}
with $\rho(s)=\sqrt{1-4m_\pi^2/s}\equiv
p/\sqrt{s}$, where $p$ is the center-of-mass momentum when $m_0^2 \to s_0$. Obviously,
the number of subtractions in Eq.~(\ref{eq:self-disp}) depends on the
assumed off-shellness. For the previous choice of $ \Gamma(s) $, which
becomes a constant at large $s$, we need at least one subtraction,
which we may choose to be, e.g., at $s=0$ and thus in terms of the
principal value integral we have
\begin{eqnarray}
\Delta m_R &=& \frac1{2m_0} \left[ 
{\rm Re} \Sigma(0)  
+ \frac1{\pi} \dashint_{4 m_\pi^2}^\infty ds' \frac{m_0^2}{s'}
\frac{{\rm Im} \Sigma(s')}{s'-m_0^2}
\right] .
\end{eqnarray}
Therefore $\Delta m_R$ depends on an arbitrary constant ${\rm Re}
\Sigma(0) = {\cal O} (N_c^{-1})$, which cannot be determined from the
dispersion integral or the lowest-order parameters and hence naively
becomes independent of the width. Other momentum-dependent widths, not
vanishing at high $s$, may introduce additional subtractions. The present
discussion illustrates our statement that one cannot generically
compute the mass shift in a model-independent way\footnote{This is so
provided no further information is available.}.

This lack of predictive power within the purely hadronic theory is not
surprising. However, from a microscopic point of view the meson
self-energy can be understood as the coupling of the $q \bar q$ bound
state to the meson continuum and physical resonances turn into
Feschbach resonances. The relevant scale corresponds to the string
breaking distance, defining a physical momentum scale which may be
described as a transition form factor from $\bar q q $ states to
mesonic channels. This implies that the mass shift due to closed
channels is necessarily negative as it corresponds to second-order
perturbation theory below the closed channels, but also that the mass
shift due to the open channel, scales exactly as the decay width. In
Appendix~\ref{sec:cut-off} we analyze some specific models where we
can see that within uncertainties a natural rough estimate of the
mass-shift is given by the half-width rule.

\subsection{Finite width effects in the space-like region}

\begin{figure*}[tb]
\includegraphics[width=7cm]{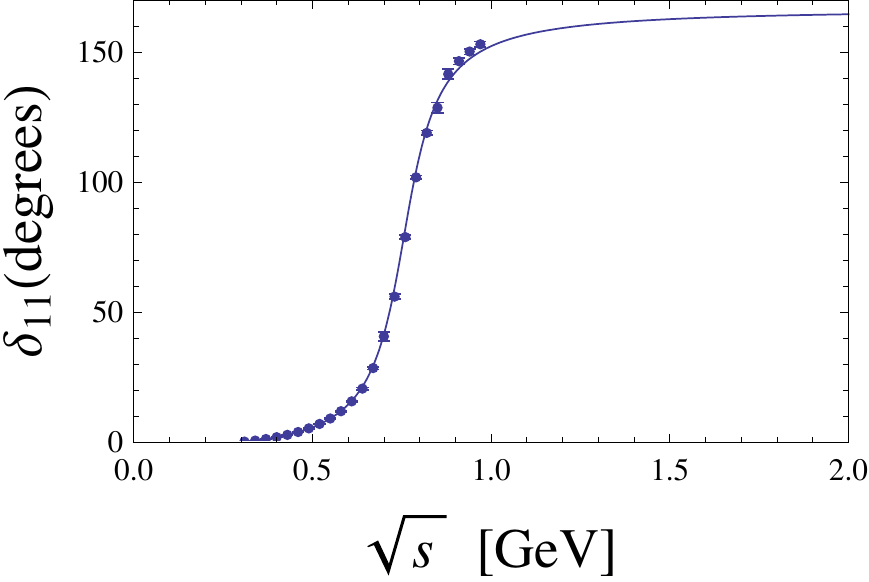}
\includegraphics[width=7cm]{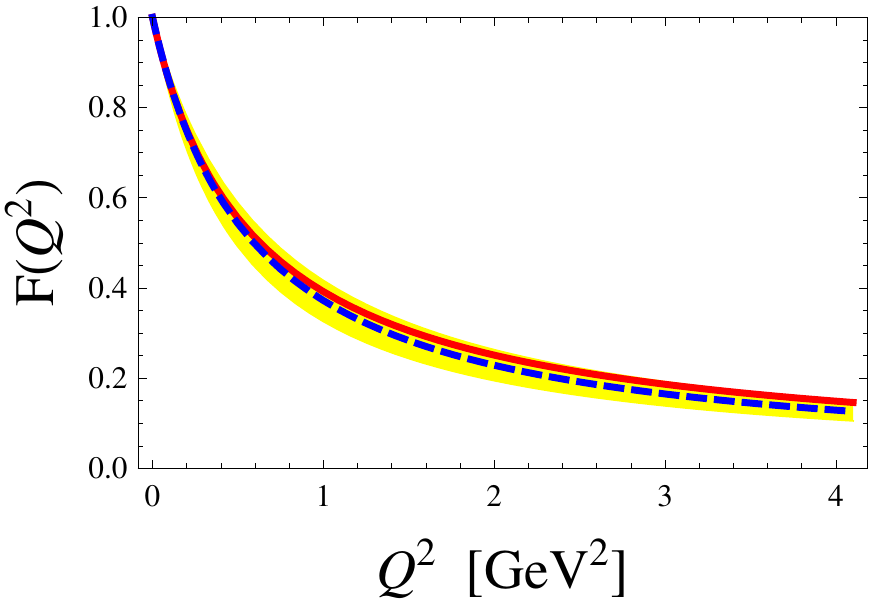}
\caption{(color online). The $(I,J)=(1,1)$ $\pi\pi$ scattering  phase shift as a
  function of the center-of-mass energy (in GeV). We use the BW representation
  discussed in the text. The data are from the analysis of
  Ref.~\cite{GarciaMartin:2011cn} (left panel). The pion charge form factor in
  the space-like region as a function of the momentum $Q^2$ (in GeV$^2$) for
  the Omnes representation (solid red), compared with the simple monopole form
  (dashed blue). The band corresponds to taking a monopole with the half-width
  rule (right panel).
\label{fig:omnes}}
\end{figure*}

Finite width corrections for the pion charge form factor were
pioneered by Gounaris and Sakurai~\cite{Gounaris:1968mw}. They have
implemented the $e^+e^- \to \pi^+ \pi^-$ final-state interactions in
the timelike region, where they are crucial. In this section we
analyze the influence of widths on the space-like region. To this end,
we use Watson's theorem on final states, which can be written
as~\footnote{Another way of writing the relation, which generalizes
  trivially to coupled channels, is through the use of the
  Bethe-Salpeter equation $ F T^{-1} = \Gamma V$~\cite{Nieves:1999bx},
  which yields ${\rm Disc} F_V(s+ {\rm i} 0^+)^{-1}={\rm Disc}
  T_{11}(s+ {\rm i} 0^+)^{-1} $. Then
$$
{\rm Disc} F_i (s+ {\rm i} 0^+)^{-1}={\rm Disc}
  \sum_j T_{ij}(s+ {\rm i} 0^+)^{-1} .
$$
Above the $\bar K K$ production threshold
Watson's theorem requires considering also the Kaon form factor and
the corresponding extension to coupled channels, i.e., the mixing with
$\pi\pi \to \bar K K$ transitions.} 
\begin{eqnarray}
\frac{F_V (s+ {\rm i} 0^+)}{F_V(s- {\rm i} 0^+)}= \frac{T_{11} (s- {\rm i} 0^+)}{T_{11} (s+ {\rm i} 0^+)} \equiv e^{2 {\rm i} \delta_{11}(s)}, \; 4m_\pi^2 \le s \le 4m_K^2,
\nonumber \\
\label{eq:watson} 
\end{eqnarray}
where $F_V(s)$ is the form factor, $T_{11} (s \pm {\rm i} 0^+)$ is the $\pi\pi$ partial-wave
scattering amplitude in the vector-isovector channel $(J,I)=(1,1)$, and
$\delta_{11}(s) $ is the corresponding phase-shift. A well known solution
to this discontinuity equation is given in terms of the Omnes function, 
\begin{eqnarray}
\Omega(s) = \exp \left[ \frac{s}{\pi} \int_{4 m^2}^{\infty}
  \frac{\delta_J(s')}{s'(s-s')} \right],
\end{eqnarray}
which fulfills $\Omega(0)=1$. A solution to Eq.~(\ref{eq:watson}) is
given by just taking $F_V(s)=P(s) \Omega(s)$, with $P(s)$ being an arbitrary
polynomial. Choosing $P(s)=1$ we have 
\begin{eqnarray}
F_V(s)=\Omega(s) .
\end{eqnarray}
In the case of a zero-width resonance the pahse-shift is  $\delta_{11}(s)= \pi/2 \, \theta(s-m_\rho^2)$ and one gets the monopole form factor, 
\begin{eqnarray}
F_{V}(s)=\frac{m_V^2}{m_V^2-s},
\label{eq:ff_v}
\end{eqnarray}
featuring VMD in its simplest version.

We use a simple Breit-Wigner (BW)
parameterization for the vector-isovector $\pi\pi$-phase shift, obtained as
\begin{eqnarray}
e^{2 i \delta_{11}(s)}=\frac{D_V (s+i0^+)}{D_{V}(s-i0^+)}\, ,
\end{eqnarray} 
where 
\begin{eqnarray}
[D_V (s)]^{-1} = s- m_\rho^2+ i m_\rho \Gamma_\rho \left[\frac{(s-4
      m_\pi^2)m_\rho^2}{(m_\rho^2-4 m_\pi^2)s}\right]^{\frac32}.
\end{eqnarray} 
The $p$-wave character of the $\rho \to 2 \pi $ decay can be
recognized in the phase space factor. The Omnes form factor is
depicted in Fig.~\ref{fig:omnes}. As we can see, the finite width
correction lies within the band corresponding to the half-width rule imposed on top of the
monopole form factor, considering here $m_\rho=0.77$~GeV and $\Gamma_\rho=0.15$~GeV.

\begin{figure*}[tbc]
%\begin{center}
\includegraphics[width=7cm]{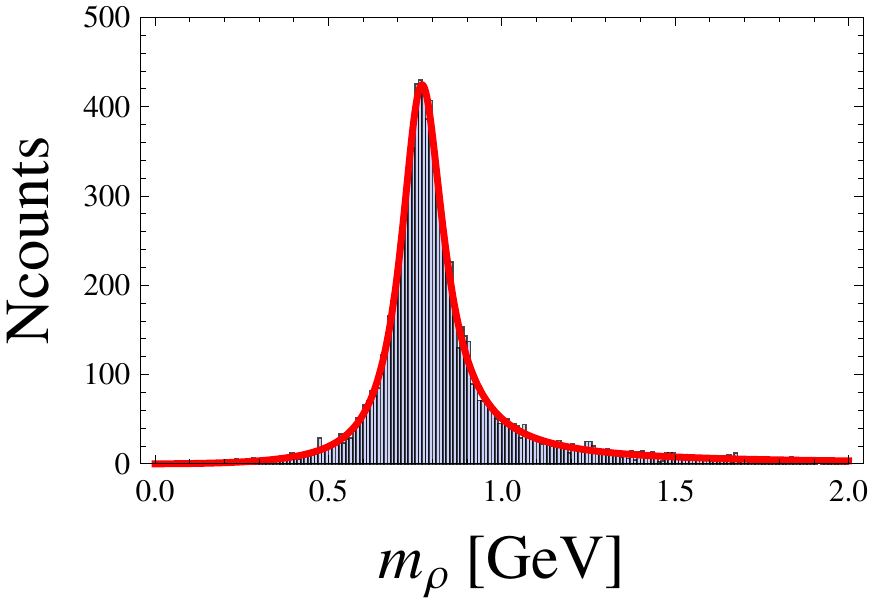} 
\includegraphics[width=7cm]{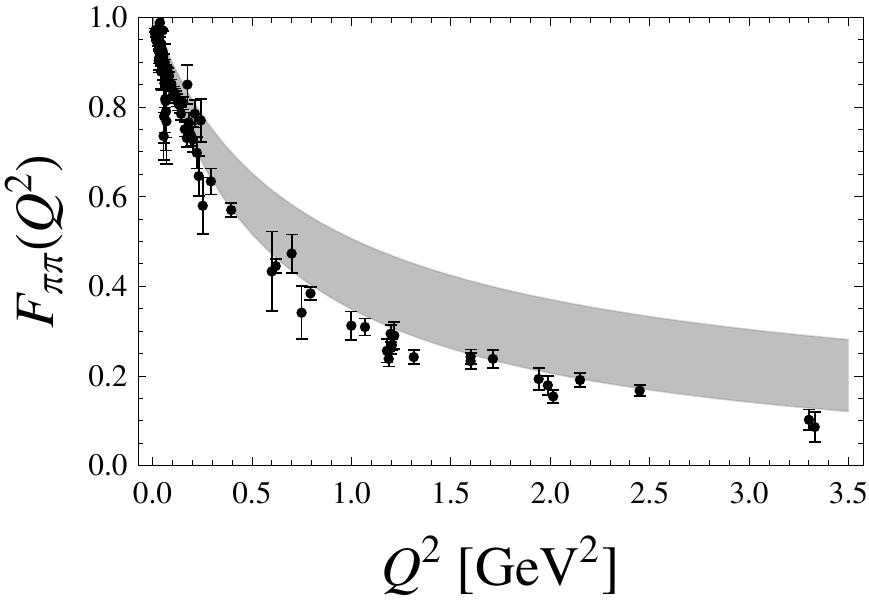}
%\end{center}
%\vspace{-3mm}
\caption{(color online). Sampling of the $\rho$-meson mass according to the
  BW spectral distribution. We sample $N=10^4$ values and bin them with
  $\Delta m= 20~{\rm MeV}$ (left panel). The monopole form factor
  $F_{\pi\pi} (Q^2)= m_\rho^2/(m_\rho^2+Q^2)$ sampled with
  the previous distribution and compared to the experimental data \cite{Brown:1973wr,Bebek:1974iz,Bebek:1974ww,Bebek:1977pe,Dally:1977vt,Brauel:1979zk,Amendolia:1986wj,Volmer:2000ek,Tadevosyan:2007yd,Horn:2006tm,Horn:2007ug} (right panel).}
\label{fig:hw-dist}
\end{figure*}

\begin{figure*}[tbc]
\includegraphics[width=7cm]{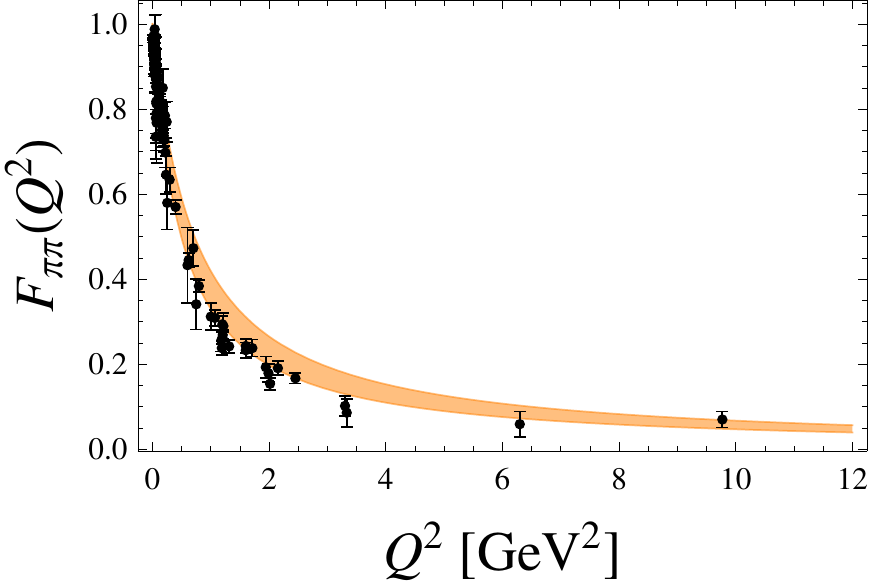}
\includegraphics[width=7cm]{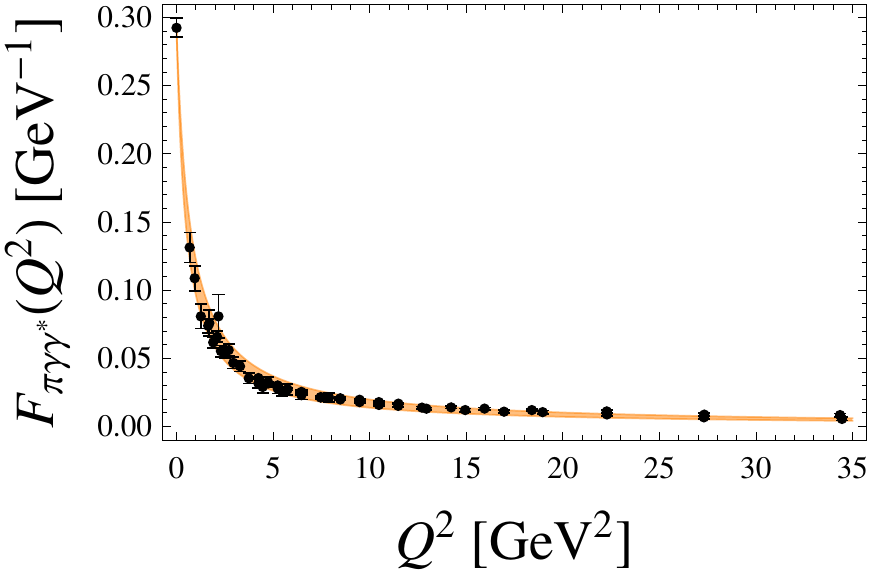}
\includegraphics[width=7cm]{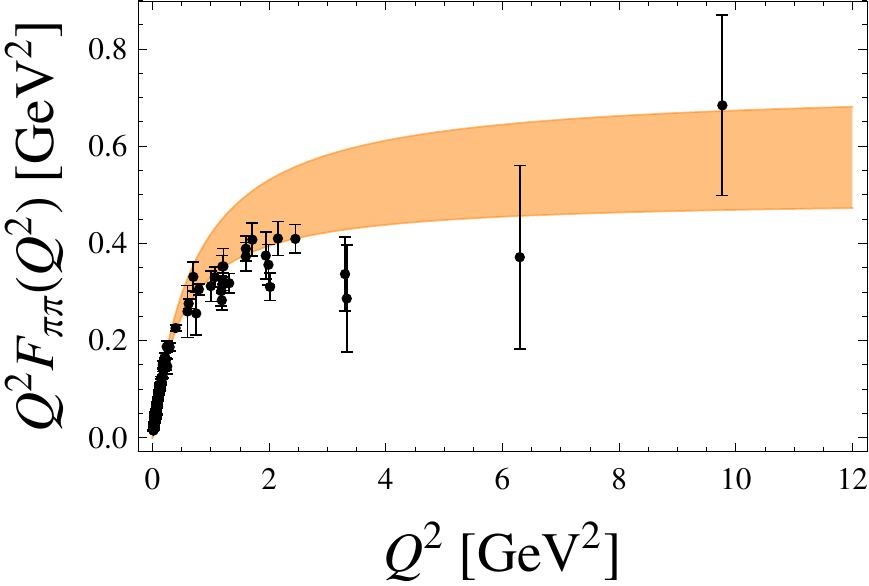}
\includegraphics[width=7cm]{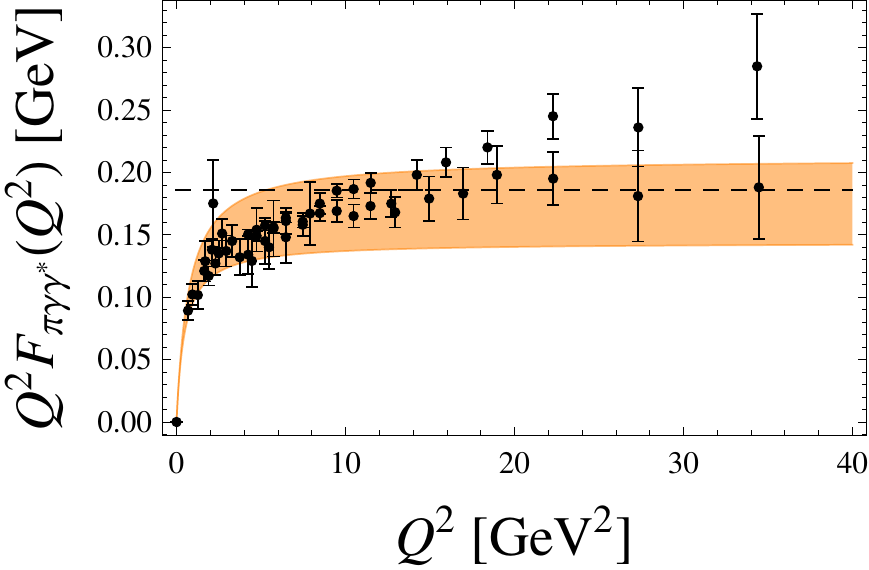}
\caption{(color online). Top row: The pion charge form factor with the half-width-rule compared with the experimental data \cite{Brown:1973wr,Bebek:1974iz,Bebek:1974ww,Bebek:1977pe,Dally:1977vt,Brauel:1979zk,Amendolia:1986wj,Volmer:2000ek,Tadevosyan:2007yd,Horn:2006tm,Horn:2007ug} (left) and the pion-photon transition form factor compared with the experimental data \cite{Behrend:1990sr,Gronberg:1997fj,Aubert:2009mc,Uehara:2012ag} (right). Bottom row: The pion charge form factor multiplied by $Q^2$ (left) and the pion-photon transition form factor multiplied by $Q^2$ (right); the horizontal dashed line represents the asymptotic value $2f_{\pi}$.
\label{fig:ftpQCD2}}
\end{figure*}

\begin{figure*}[tbc]
\includegraphics[width=7cm]{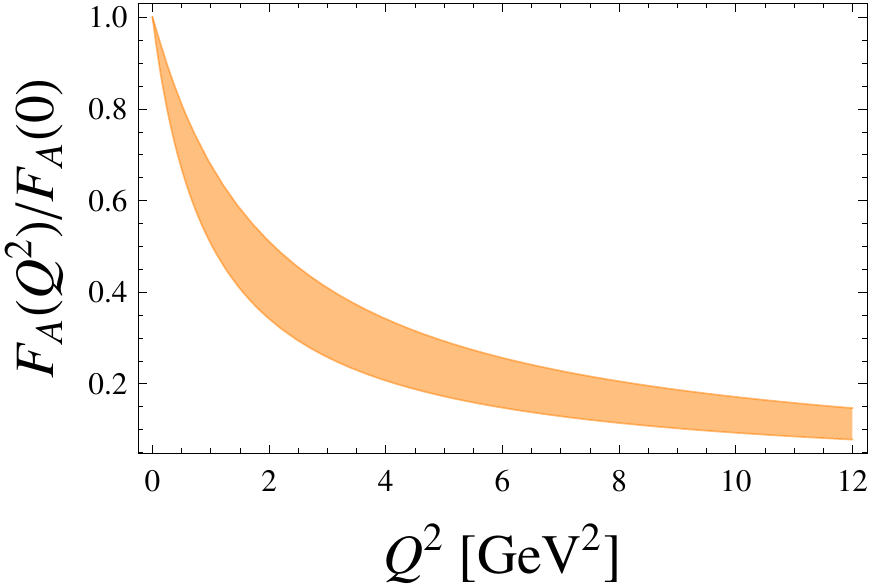}
\includegraphics[width=7cm]{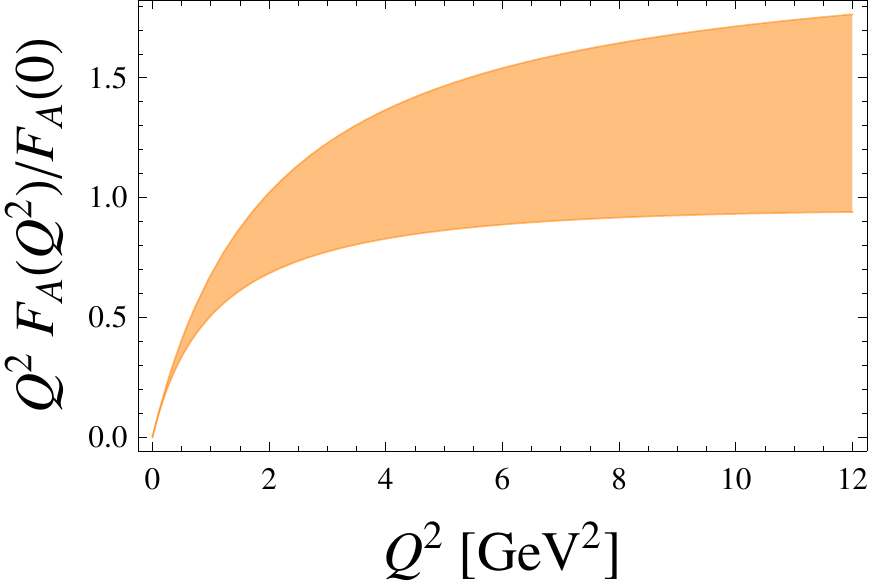}
\caption{(color online). Axial form factor
in the space-like region $t <0$
\label{fig:ftpQCD3}}
\end{figure*}

\subsection{The half-width rule}

As we see, the subleading $1/N_c$ corrections in the space-like region
essentially correspond to keeping the meson dominance form and to changing the
parameters. 
%
% Too many caveats!
%
%On the other hand, a reliable calculation of these
%modifications is quite cumbersome and it is unclear whether it is
%complete from the point of view of the effective theory if just a
%finite number of mesons is kept. 
By making simple calculations we have
seen that a conservative bound on the mass shift is given by the
half-width rule. From a spectral point of view this is quite natural when we appeal to 
the K\"all\'en-Lehmann representation of the resonance two-point function, 
\begin{eqnarray}
D(s)= \int_0^\infty d \mu^2 \frac{\rho(\mu^2)}{\mu^2-s-i {0^+}} .
\end{eqnarray}
We may then use the probabilistic interpretation of the line shape
\begin{eqnarray}
P(\mu)= Z \rho(\mu) .
\end{eqnarray}
If we take a Breit-Wigner shape for $D(s)$ (neglecting the threshold
effects), we have
\begin{eqnarray}
P_{\rm BW}(\mu)= \frac1{\pi}\frac{2 \Gamma \mu^2}{(\mu^2-M^2)^2 +\Gamma^2 \mu^2},
\end{eqnarray}
which is normalized to unity, $\int d\mu P(\mu)=1$. 

The random
implementation for a given distribution is obtained in a standard way 
by inverting the relation expressing the coordinate independence of probabilities,
\begin{eqnarray}
P(\mu) d\mu  = dz, 
\end{eqnarray}
with $z \in U[0,1]$ being a uniformly distributed
variable\footnote{For a Lorentz distribution, $P(\mu) = Z \mu /(
  (\mu^2-M_R^2)^2+ \Gamma_R^2 M_R^2)$, the relation is given by $
  \mu^2 = M_R^2 + M_R \Gamma_R \tan\left(\pi z/2\right) $, with $ z
  \in U[0,1] $. This distribution must be cut and normalized when
  negative $\mu^2$ values are generated.}. The result for a BW shape
in the case of the $\rho$-meson for $N=10^4$ samples is shown in
Fig.~\ref{fig:hw-dist}.  The idea amounts to treating the resonance
mass as a random variable and to propagating its effect in all
observables. Of course, different shapes produce somewhat different
confidence levels. For definiteness, we will take Gaussians which have
shorter tails and are symmetric around the resonance value.

In Fig.~\ref{fig:hw-dist} we plot the monopole form factor 
\begin{eqnarray}
F_V(Q^2) = 
\frac{m_\rho^2}{m_\rho^2+Q^2}
\end{eqnarray}
according to the BW distribution of the mass.  

%The mass-shift of a resonance which becomes stable in the large
%$N_c$-limit satisfies a dispersion relation which 

\section{Pion form factors}
\label{sec:pion-FF}

\subsection{Electromagnetic form factor}\label{subsec:EMFF}

The charge form factor of the pion is given by 
\begin{eqnarray}
\langle \pi^+ (p') | J_\mu^{\rm em} (0) | \pi^+ (p) \rangle = 
\left(p'^\mu + p^\mu \right) F_V(q^2),     
\end{eqnarray} 
with $q=p'-p$ and the electromagnetic current $ J_\mu^{\rm em} (x) = \sum_{q=u,d,s, \dots} e_q\bar
q (x) \gamma_\mu q(x)$, where $e_q$ 
denotes the quark charges in units of the elementary charge. 
The charge normalization requires
\begin{eqnarray}
F_V(0)= 1.
\label{eq:FV0}
\end{eqnarray} 
Actually, in the space-like region, where $t=-Q^2$, $F(t)$ is real and
at large $Q^2$ values the pQCD methods can be applied, yielding
asymptotically~\cite{Brodsky:1973kr,Brodsky:1974vy,Farrar:1979aw,Radyushkin:1977gp,Efremov:1978rn,Efremov:1979qk}
\begin{eqnarray}
F_V(-Q^2) &=& \frac{16 \pi f_\pi^2 \alpha(Q^2)}{Q^2}\left[1+ 6.58
  \frac{\alpha (Q^2)}{\pi} + \dots \right], \nonumber 
\\ \qquad && Q^2 \gg m^2 \label{pQCDff}
\end{eqnarray} 
with $f_\pi=92.3$~MeV denoting the pion weak decay constant, and
$m$ standing for the lowest vector meson mass. If we ignore the slowly varying
logarithm, we get $F_V(t) = {\cal O}(t^{-1})$ and in the large $N_c$
limit one has
\begin{eqnarray}
F_V(t) = 
\sum_{V=\rho,\rho', \dots} c_V\frac{m_V^2}{m_V^2-t},
\end{eqnarray}
where $\sum_V c_V=1$ and $c_V = g_{V\pi\pi} F_V /m_V$. 

The simplest formula fulfilling 
this constraint and (\ref{eq:FV0}) is the VMD solution 
\begin{eqnarray}\label{VMDeq}
F_V(t) = 
\frac{m_\rho^2}{m_\rho^2-t},
\end{eqnarray}
whence $g_{\rho\pi\pi} f_\rho= m_\rho$. 
The $\rho-\gamma$ coupling is given by
\begin{eqnarray}\label{eq:rhotoee}
\Gamma( \rho \to e^+ e^-) = \frac{4 \pi \alpha^2}{3} \frac{F_\rho^2}{m_\rho},
\end{eqnarray}
whereas the $\rho \to \pi\pi$ decay is 
\begin{eqnarray}
\Gamma( \rho \to \pi \pi) = \frac{g_{\rho \pi \pi}^2 m_\rho}{48 \pi}.
\end{eqnarray}
Using the PDG numbers one gets for $m_\rho=0.77~{\rm GeV}$ 
the values $f_\rho = 0.156~{\rm GeV}$ and $g_{\rho\pi\pi} = 6$.

One of the important features of the pion form factor is that the
radius has large chiral corrections, thus we may improve the
phenomenology by adding one extra meson, $\rho'$. Then
\begin{eqnarray}\label{extVMDeq}
F_V(t) = (1-c)\frac{m_\rho^2}{m_\rho^2-t}+ c\frac{m_{\rho'}^2}{m_{\rho'}^2-t},
\end{eqnarray}
such that 
\begin{eqnarray}
\frac16\langle r^2 \rangle  = (1-c)\frac{1}{m_\rho^2}+ c\frac{1}{m_{\rho'}^2}.
\end{eqnarray}
Thus imposing the physical value of $\langle r^2 \rangle $ we get the value of
$c$ for any $m_\rho$ and $m_{\rho'}$. Taking again the PDG values for those quantities ($m_\rho=0.77549(34)$~GeV, $m_\rho'=1.465(25)$~GeV, 
$\langle r^2 \rangle=(0.672(8)\textrm{fm})^2 $) we obtain $c=-0.227(39)$.

One could also carry out the analysis the other way around, starting 
from Eqs.~(\ref{VMDeq},\ref{extVMDeq}) with all the 
constants treated as free parameters to be determined by a fit to the experimental data. That way it 
was shown in Refs.~\cite{Masjuan:2008fv,Queralt:2010sv} that one can retain a precise value for $\langle r^2 \rangle $, 
even though the masses do not have precisely their physical values. In the large-$N_c$ limit, when the functions become 
meromorphic, this fitting procedure is mathematically safe thanks to the convergence theorems from 
the Pad\'e Theory \cite{Queralt:2010sv}. In this framework, Fig.~3 of Ref.~\cite{Masjuan:2008fv} shows the half-width-rule 
as a good estimation of the systematic error done on the determination of the poles when fitting the space-like data.

\subsection{Axial form factor}

The axial form factor of the pion is intimately related to the pion radiative decay $\pi^{\pm} \to l^{\pm} \gamma \nu$ (with $l$ standing for $e$ or $\mu$) and its hadronic contribution. The decay proceeds via ordinary inner bremsstrahlung (IB) from the weak decay $\pi^{\pm}\rightarrow l^{\pm}\nu$ accompanied by the photon radiated from the external charged particles, and the structure-dependent interaction (SD) between the photon and the virtual hadronic states, with contributions of both vector and axial-vector form factors.

For the transition $\pi^+ \to \gamma \nu_e e^+
$~\cite{Bryman:1982et}, the structure dependent
amplitude is given by 
\begin{eqnarray}
M_{\rm SD} = i e G_F \cos \theta_c \bar u_\nu
\gamma_\mu ( 1 - \gamma_5 ) v_e \epsilon^*_\nu M^{\mu \nu } / \sqrt{2}
m_\pi, 
\end{eqnarray}
where $\epsilon^{\mu}$ is the polarization vector of the photon, $G_F$ is the weak interaction coupling constant, and $\theta_c$ is the Cabibbo angle.
The hadronic contribution is enclosed in the amplitude $M^{\mu \nu }$ and reads:
\begin{eqnarray}
&& M^{\mu \nu} =
F_V(t) \epsilon^{\mu \nu \alpha \beta} p_\alpha q_\beta \nonumber \\
&& -i F_A(t) [
  q^{\nu} (q^\mu + p^\mu) - g^{\mu \nu} q \cdot (q+p) ]\, ,
\end{eqnarray}
with $F_{V,A}(t)$ denoting the vector and axial-vector form factors, respectively. By assuming axial-meson dominance, we have 
\begin{eqnarray}
F_A(t)=F_A(0) \frac{M_A^2}{M_A^2-t}\,
\end{eqnarray}
with $F_A(0)=0.0119(1)$ \cite{PDG2012}. 
The result obtained with the half-width-rule is presented in Fig.~\ref{fig:ftpQCD3}.

\subsection{Transition form factor}

The pion-photon transition form factor $\pi^0 \to \gamma \gamma^*$ has been
subjected to vigorous discussion in recent years. Firstly, its value at the origin is
fixed by the chiral anomaly,
\begin{eqnarray}
F(0)= \frac1{4 \pi^2 f_\pi},
\end{eqnarray}
while its asymptotic behavior is given by 
\begin{eqnarray}
F(Q^2) \to  \frac{6 f_\pi}{ N_c Q^2}+ \dots .
\end{eqnarray}
A simple model fulfilling both conditions is 
\begin{eqnarray}
F(Q^2)= \frac1{4 \pi^2 f_\pi}\frac{m_\rho^2}{m_\rho^2+Q^2} ,
\end{eqnarray}
provided one has the relation
\begin{eqnarray}
m_\rho^2 = \frac{24 \pi^2 f_\pi^2}{N_c},
\end{eqnarray}
which gives $m_\rho=823$~MeV for $f_\pi = 92.6$~MeV
or $m_\rho=770$~MeV for $f_\pi=86.6$~MeV in the chiral limit. 

\begin{figure}[tbc]
\includegraphics[width=7cm]{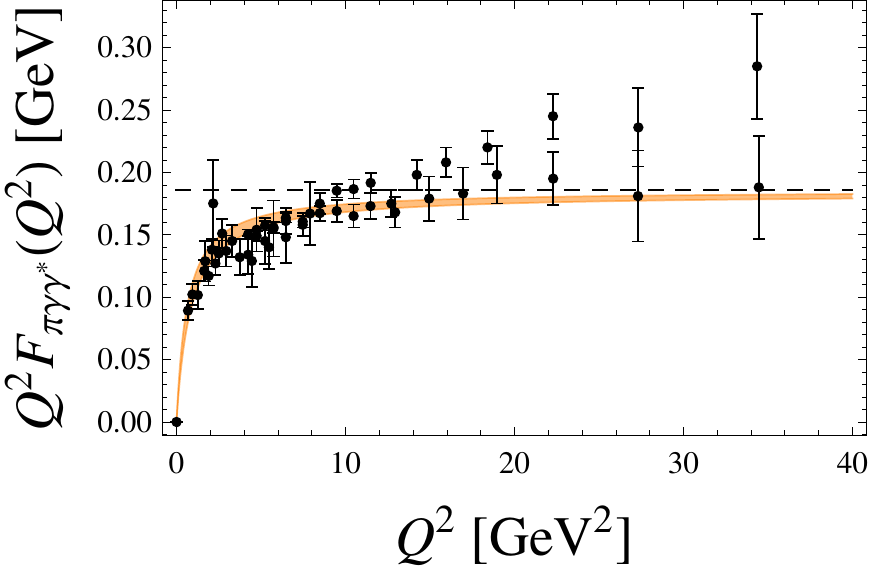}
\caption{(color online). Band: the pion-photon transition form factor of Eq.~(\ref{eq:2m}). % with $m_\rho=770 {\rm ~MeV}$ and $m_\rho'=1465 {\rm ~MeV}$.
Points: various experimental data \cite{Behrend:1990sr,Gronberg:1997fj,Aubert:2009mc,Uehara:2012ag}. The horizontal line represents the 
theoretic asymptotic value of $ 2 f_\pi$. 
\label{fig:ftpQCD}}
\end{figure}

If we include two resonances \cite{Knecht:2001qf}, $\rho$ and $\rho'$, we get, 
after imposing the anomaly and large-$Q^2$ behavior,  
\begin{eqnarray}
F(Q^2)= \frac1{4 \pi^2 f_\pi}\frac{m_\rho^2 m_{\rho'}^2 + 24 f_\pi^2
  \pi^2 Q^2 /N_c }{(m_\rho^2+Q^2)(m_{\rho'}^2+Q^2)}. \label{eq:2m}
\end{eqnarray}
The result is shown in Fig.~\ref{fig:ftpQCD}, using $m_\rho=0.775$~GeV, $m_\rho'=1.465$~GeV, $\Gamma_\rho=0.150$~GeV and $\Gamma_\rho'=0.400$~GeV. 

One could even go beyond this approximation including a third resonance (the $\rho''$). That introduces a new parameter that 
can be fixed by the derivative of the form factor at the origin, the parameter $a_{\pi}$ \cite{Masjuan:2012wy}:
\begin{eqnarray}
&&F(Q^2)=\\&& \nonumber
\frac1{4 \pi^2 f_\pi}\frac{m_\rho^2 m_{\rho'}^2 m_{\rho''}^2 + \textrm{b}\, Q^2 +24 f_\pi^2
  \pi^2 Q^4 /N_c }{(m_\rho^2+Q^2)(m_{\rho'}^2+Q^2) (m_{\rho''}^2+Q^2)}\, ,\label{eq:3m}
\end{eqnarray}
where the parameter $b$ can be obtained through a matching procedure to the low-energy expansion of $F(Q^2)$, i.e., 
$b=m_{\rho}^2 m_{\rho'}^2 m_{\rho''}^2 (\frac{a_{\pi}}{m_{\pi}^2} + \frac1{m_{\rho}^2}+\frac1{m_{\rho'}^2}+\frac1{m_{\rho''}^2})$. 
Given $m_\pi=0.135$~GeV, $m_\rho, m_\rho'$, $m_\rho''=1.720(20)$~GeV, and  $a_\pi=0.032(4)$, we obtain $b=5.82(18)$.

Figure~3 of Ref.~\cite{Masjuan:2012wy} shows how the half-width-rule provides a good estimate of the 
systematic error on the determination of poles of rational approximants, such as Eqs.~(\ref{eq:2m},\ref{eq:3m}), when fitting to the space-like data \cite{Behrend:1990sr,Gronberg:1997fj,Aubert:2009mc,Uehara:2012ag}.

\subsection{Gravitational form factor}

The gravitational quark form factors of the pion
\cite{Donoghue:1991qv}, $\Theta_1$ and $\Theta_2$, are defined through
the matrix element of the quark part of the energy-momentum tensor in the one-pion
state,
\begin{eqnarray}\label{eq:gravFF}
&& \!\!\!\!\!\! \langle \pi^b(p') \mid \Theta^{\mu \nu}(0) \mid \pi^a(p) \rangle = \\ && 
\!\!\!\!\!\!\! \frac{1}{2}{\delta^{ab}}\left [ (g^{\mu \nu}q^2- q^\mu q^\nu)
\Theta_1(q^2)+ 4 P^\mu P^\nu \Theta_2(q^2) \right ], \nonumber
\end{eqnarray}
where $P=\frac{1}{2}(p'+p)$, $q=p'-p$, and $a, b$ are the isospin
indices.  The gravitational form factors satisfy the low-energy
theorem $\Theta_1(0) - \Theta_2(0) = {\cal O} (m_\pi^2)
$~\cite{Donoghue:1991qv}. 
The trace part is dominated by scalar states, while the traceless component is dominated by spin-2 tensor mesons.  

Following the standard notation for the moments of the pion generalized parton distributions (GPDs) \cite{Broniowski:2009zh}, we introduce:
\begin{equation}
A_{20}(t)=\frac12\Theta_1(t)\, ,\quad\quad A_{22}(t)=-\frac12\Theta_2(t)\, ,
\end{equation}
where the symbols $\Theta_i(t)$ denote the quark parts of the gravitational form factors of Eq. (\ref{eq:gravFF}). We consider also the moment
\begin{equation}
A_{10}(t)=F_V(t)\, ,
\end{equation}
with $F_V(t)$ denoting the electromagnetic form factor described in Sec. \ref{subsec:EMFF}.

\begin{figure*}[htbp]
\begin{center}
\includegraphics[width=7cm]{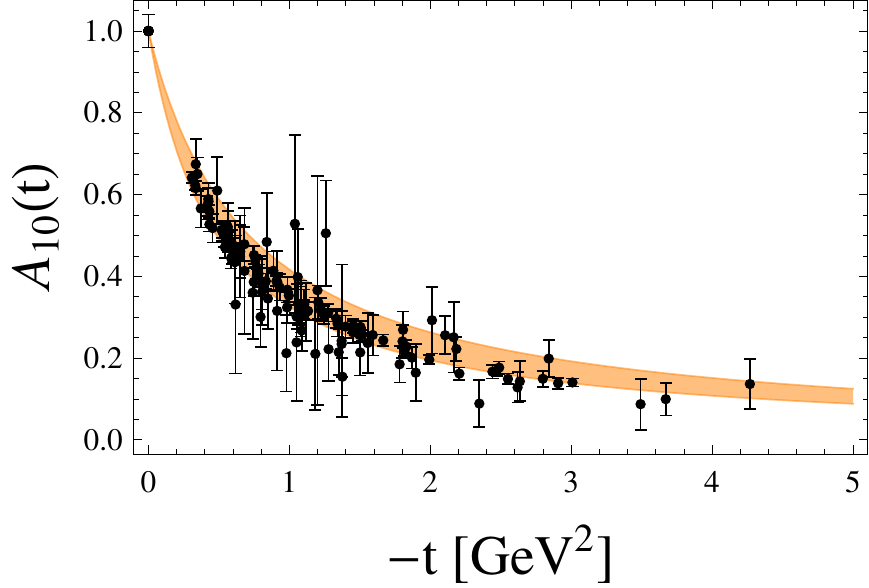}
\includegraphics[width=7cm]{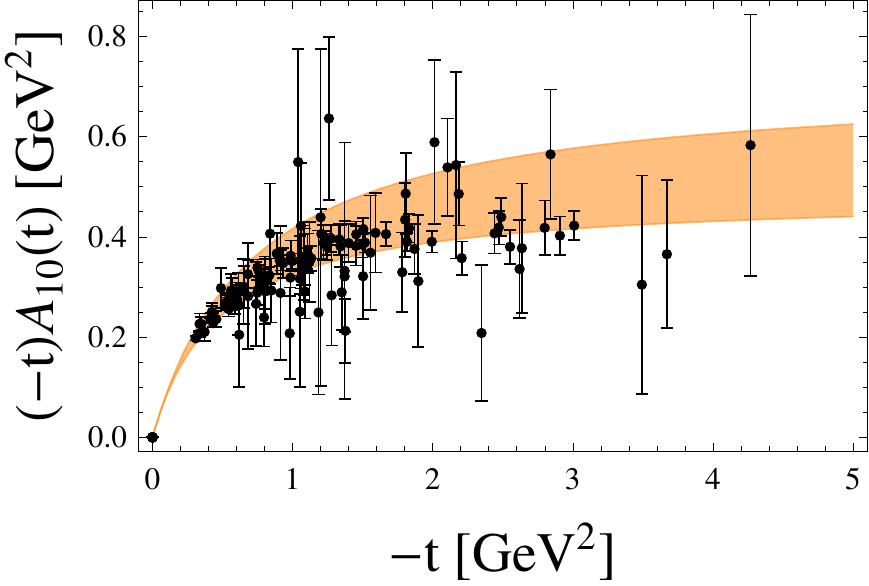}
\includegraphics[width=7cm]{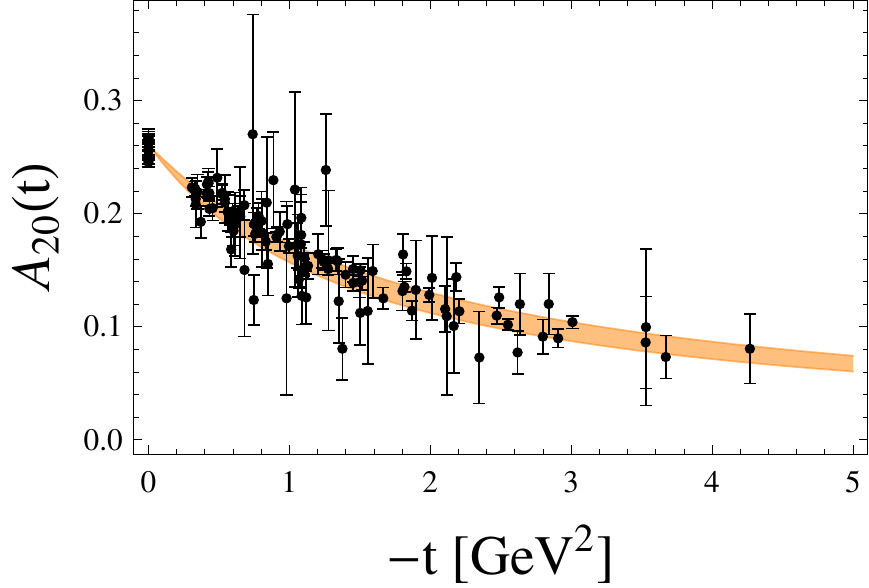}
\includegraphics[width=7cm]{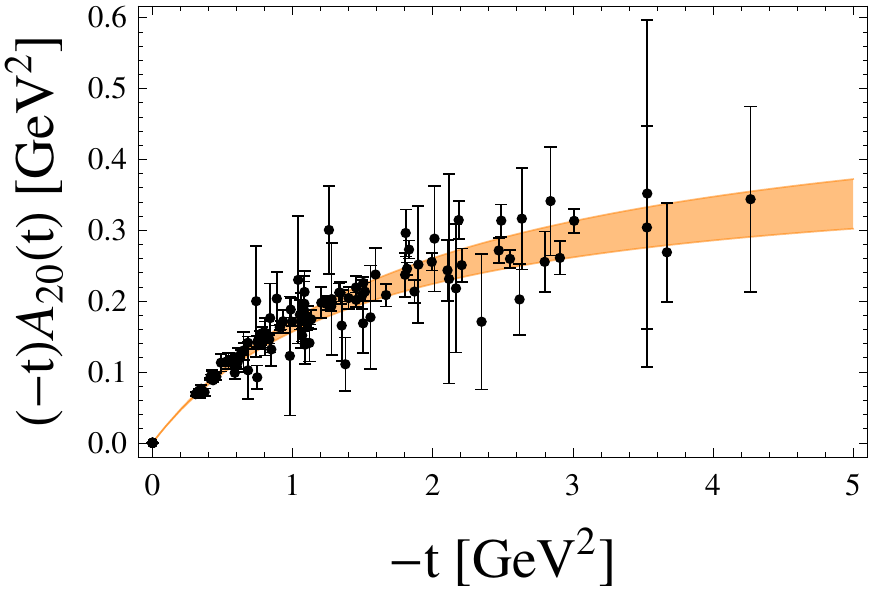}
\includegraphics[width=7cm]{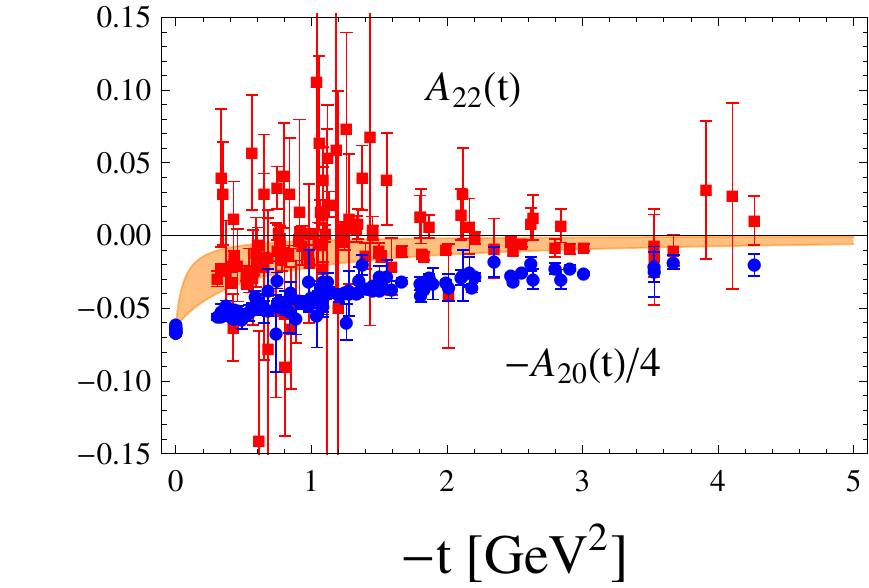}
\caption{(color online). Electromagnetic form factor $A_{10}(t)$, the corresponding $(-t) A_{10}(t)$,  and 
the quark part of the gravitational form factor $A_{20}(t)$ and $(-t)A_{20}(t)$, compared to the lattice data from \cite{Brommel:PhD}  (top and middle row respectively).  Spin-0 gravitational form factor of the pion, $A_{22}(t)$, from the lattice calculation of Refs. \cite{Brommel:PhD,Brommel:2007xd}  extrapolated to the physical pion mass. Red squares correspond to $A_{22}(t)$ and blue circles to $-A_{20}(t)/4$ (bottom row).}
\label{a10plot}
\end{center}
\end{figure*}

The results of monopole representations with the half-width rule are shown in Fig.\ref{a10plot}, where we consider
\begin{equation}
A_{10}(t)=\frac{m_\rho^2}{m_\rho^2-t}\, , \quad \quad A_{20}(t)=A_{20}(0) \frac{m_{f_2}^2}{m_{f_2}^2-t}\,,
\end{equation}
with $m_{f_2}=1.320$~GeV, $\Gamma_{f_2}=0.185~$GeV, and $A_{20}(0)=0.261$.
We also show in Fig.~\ref{a10plot} the low-energy theorem that relates $A_{20}(t)$ with the other moment, $A_{22}(t)$. 
The relation $A_{22}(0)=-\frac14A_{20}(0)$ is compared to the lattice data of Refs.~\cite{Brommel:PhD,Brommel:2007xd}.

%\begin{figure}[htbp]
%\begin{center}
%\includegraphics[width=7cm]{A10tplotExp.pdf}
%\caption{{OPTIONAL FIGURE, IT MIGHT BE INTERCHANGED WITH UPPER RIGHT PLOT ON FIG.6}. Electromagnetic for factor $(-t) A_{10}(t)$ compared to both lattice data from Ref.\cite{Brommel:PhD} (black circles) and experimental data from  Refs. \cite{Brown:1973wr,Bebek:1974iz,Bebek:1974ww,Bebek:1977pe,Dally:1977vt,Brauel:1979zk,Amendolia:1986wj,Volmer:2000ek,Tadevosyan:2007yd,Horn:2006tm,Horn:2007ug} (red squares).  }
%\label{a10plotbis}
%\end{center}
%\end{figure}

\section{Nucleon Form Factors}
\label{sec:nuclen-FF}

\subsection{Electromagnetic form factors}

In the non-strange sector the electromagnetic current is given by 
\begin{eqnarray}
J^\mu_{\rm em} (x) = \frac12 J_B^\mu(x) + J_V^{\mu 3} (x)  
\end{eqnarray} 
where $J_B^\mu(x) $ is the baryon current and $J_V^{\mu 3} (x) $ is the third
component of the isospin current
\begin{eqnarray}
J_B^{\mu } (x) &=& \bar q (x) \gamma^\mu q (x), \nonumber  \\ 
J_V^{\mu a } (x) &=& \bar q (x) \gamma^\mu \frac{\tau^a}{2}q (x) . 
\end{eqnarray}  
The matrix elements of these currents are 
%\begin{widetext} 
\begin{eqnarray}
&&\langle N(p') | J_B^\mu (0) | N(p) \rangle = \nonumber \\ 
&&\bar u (p') \left[ 
\gamma^\mu F_1^{I=0} (q^2) + \frac{i \sigma^{\mu \nu} q_\nu}{2M_N}
F_2^{I=0} (q^2) \right] u (p), \nonumber \\
%\end{eqnarray} 
%\begin{eqnarray}
&&\langle N(p') | J_V^{\mu a} (0) | N(p) \rangle = \nonumber \\
&& \bar u (p')
 \frac{ \tau^a}2 \left[ \gamma^\mu F_1^{I=1} (q^2) + \frac{i
\sigma^{\mu \nu} q_\nu}{2M_N} F_2^{I=1} (q^2) \right] u (p), \nonumber \\ 
\end{eqnarray} 
%\end{widetext} 
where $q=p'-p$, and $F_1$ and $F_2$ are the Dirac and Pauli form factors, respectively.  The
relation to the proton and neutron form factors is
\begin{eqnarray} 
F_i^p &=&  ( F_i^{I=0} +F_i^{I=1} ), \nonumber  \\
F_i^n &=&  ( F_i^{I=0} -F_i^{I=1} ),
\end{eqnarray} 
where 
\begin{eqnarray}
F_1^p(0) =1, \qquad  F_1^n(0) =0, \nonumber \\
F_2^p(0) = \kappa_p,  \qquad  F_2^n(0) = \kappa_n. \\
\end{eqnarray} 
The quantities $\kappa_p = 1.793 $ and $\kappa_n = -1.913 $
are the anomalous proton and neutron magnetic moments, respectively.
The electric and magnetic Sachs form factors are defined as
\begin{eqnarray}
G_E^p (q^2) &=& F_1^p (q^2) + \frac{q^2}{4M_N^2} F_2^p (q^2), \nonumber  \\ 
G_E^n (q^2) &=& F_1^n (q^2) + \frac{q^2}{4M_N^2} F_2^n (q^2), \nonumber  \\
G_M^p (q^2) &=& F_1^p (q^2) + F_2^p (q^2), \nonumber \\ 
G_M^n (q^2) &=& F_1^n (q^2) + F_2^n (q^2).
\end{eqnarray} 
The normalization conditions become
\begin{eqnarray}
G_E^p(0) =1, \qquad  G_E^n(0) =0, \nonumber \\
G_M^p(0) = \mu_p,  \qquad  G_M^n(0) = \mu_n, \\
\end{eqnarray} 
where $\mu_p = 2.79 \mu_N $ and $\mu_n = -1.91 \mu_N $, with $\mu_N=e
/(2 M_N)$ denoting the nuclear magneton. 

The asymptotic behavior for $t \to -\infty $ is given
by~\cite{Brodsky:1976rz}
\begin{eqnarray}
t^{i+1} F_i (t) \to \left[ \log (-t/\Lambda^2) \right]^{-\gamma}, \;\; (i=1,2) 
\label{asymptotic} 
\end{eqnarray}
where the anomalous dimension $\gamma \sim 2$ is weakly depending on
the number of flavors. As mentioned before, such a slowly changing
log behavior cannot be reproduced with a finite number of resonances,
thus we assume it to be constant. At the 
leading order in the large-$N_c$ expansion the form factors read
\begin{eqnarray}
 F_1^{I=0} (t)&=& \sum_{V } \frac{g_{\omega NN} f_{\omega \gamma}}{m_\omega^2-t}, \nonumber \\ 
F_2^{I=0} (t)&=& \sum_{V } \frac{f_{\omega NN} f_{\omega \gamma}}{m_\omega^2-t}, \nonumber  \\ 
F_1^{I=1} (t) &=& \sum_{V } \frac{g_{\rho NN} f_{\rho \gamma}}{m_\rho^2-t}, \nonumber  \\ 
F_2^{I=1} (t) &=& \sum_{V } \frac{f_{\rho NN} f_{\rho \gamma }}{m_\rho^2-t}. 
\end{eqnarray} 
We define the strong isoscalar and isovector vertices 
\begin{eqnarray}
\langle N(p') | \omega^{\mu}   | N(p) \rangle &=& 
\bar u(p') \left[ g_{\omega NN} \gamma^\mu + f_{\omega NN} 
\frac{i \sigma^{\mu \nu} q_\nu}{2M_N} \right] u(p), \nonumber \\ 
\langle N(p') | \rho_a^{\mu}   | N(p) \rangle &=& 
\bar u(p') \frac{\tau_a}{2}\left[ g_{\rho NN} \gamma^\mu + f_{\rho NN} 
\frac{i \sigma^{\mu \nu} q_\nu}{2M_N} \right] u(p). \nonumber \\ 
\end{eqnarray} 
According to Eq.~(\ref{asymptotic}), the minimum number of resonances is two and three for the Dirac and 
Pauli form factors, respectively. We
use the normalization conditions at the origin, the asymptotic conditions, and fix
the vector and tensor couplings to the lowest-lying resonance,
$g_{\omega NN}$, $f_{\omega NN}$, $g_{\rho NN}$, $f_{\rho NN}$ in the isoscalar and isovector 
channels. Our illustrative goal here is to predict the form factors without attempting a detailed fit to the data.

After all the conditions are imposed, we get 
\begin{eqnarray}
F_1^{I=0}(t) &=& \frac12 \frac{1- c_0 t
  /m_{\omega''}^2}{(1-t/m_\omega^2)(1-t/m_{\omega'}^2)
  (1-t/m_{\omega''}^2)}, \nonumber \\
F_1^{I=1}(t) &=& \frac12 \frac{1- c_1 t
  /m_{\rho''}^2}{(1-t/m_\rho^2)(1-t/m_{\rho'}^2)
  (1-t/m_{\rho''}^2)}, \nonumber \\
F_2^{I=0}(t) &=& \frac12 \frac{1}{(1-t/m_\omega^2)(1-t/m_{\omega'}^2)
  (1-t/m_{\omega''}^2)}, \nonumber \\
F_2^{I=1}(t) &=& \frac12 \frac{1}{(1-t/m_\rho^2)(1-t/m_{\rho'}^2)
  (1-t/m_{\rho''}^2)}, \nonumber \\
\end{eqnarray} 
where the constants $c_0$ and $c_1$ are determined from the values of
$g_{\omega NN}$ and $g_{\rho NN}$, respectively, as 
\begin{eqnarray}
\frac{g_{\omega NN} f_{\omega \gamma } }{m_\omega^2} &=&\frac12 \frac{1- c_0 m_\omega^2 /m_{\omega''}^2}{(1-m_\omega^2/m_{\omega'}^2)(1-m_\omega^2/m_{\omega''}^2)}, \nonumber \\
\frac{g_{\rho NN} f_{\rho \gamma } }{m_\rho^2} &=&\frac12 \frac{1- c_0 m_\rho^2 /m_{\rho''}^2}{(1-m_\rho^2/m_{\rho'}^2)(1-m_\rho^2/m_{\rho''}^2)}.
\end{eqnarray} 
Thus, with the electromagnetic decay widths for the vector mesons, 
$F_{\rho } = 0.149~{\rm GeV}$, we find (Eq.(\ref{eq:rhotoee}))
$\Gamma (\rho \to e^+ e^-) = 4 \pi \alpha^2 F_\rho^2/m_\rho/3$,
which  is numerically equal to $6.4$~KeV.

Detailed fits implementing VMD~\cite{Dubnicka:2002yp} require large
OZI violations as well as huge departure from the flavor $SU(3)$ symmetry (see also \cite{Faessler:2009tn}).
In particular, the value of $g_{\omega NN} \gg g_{\omega NN}^{\rm SU(3)}$~\cite{Mergell:1995bf,Belushkin:2006qa}. 
However, the simple Kelly parameterization~\cite{Kelly:2004hm} provides a
successful fit in the space-like region as a rational function with the correct large momentum behavior.

\begin{figure*}[tbc]
\includegraphics[width=7cm]{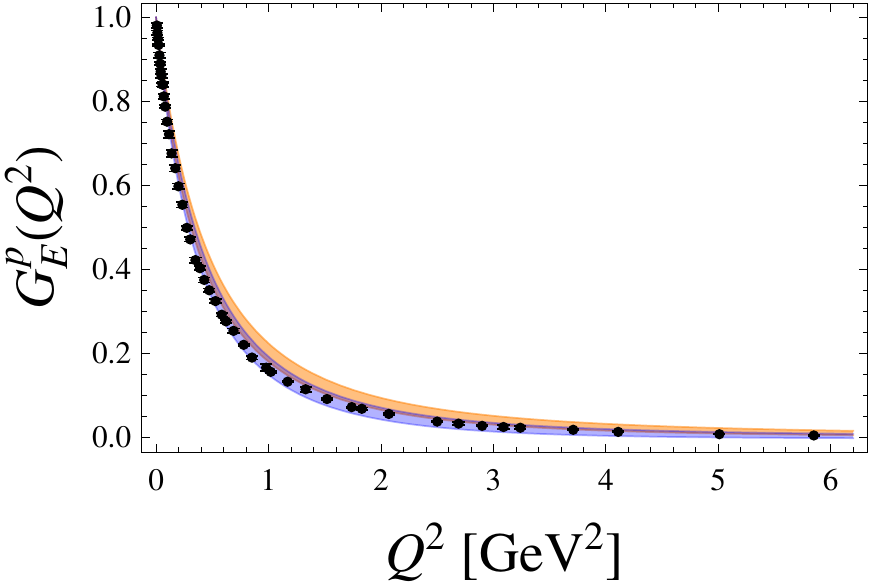}
\includegraphics[width=7cm]{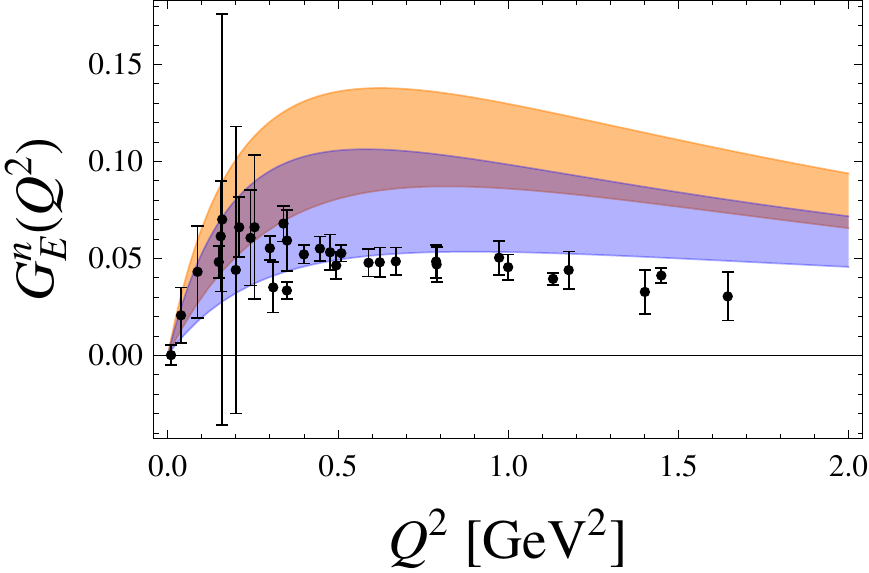} \\ 
\includegraphics[width=7cm]{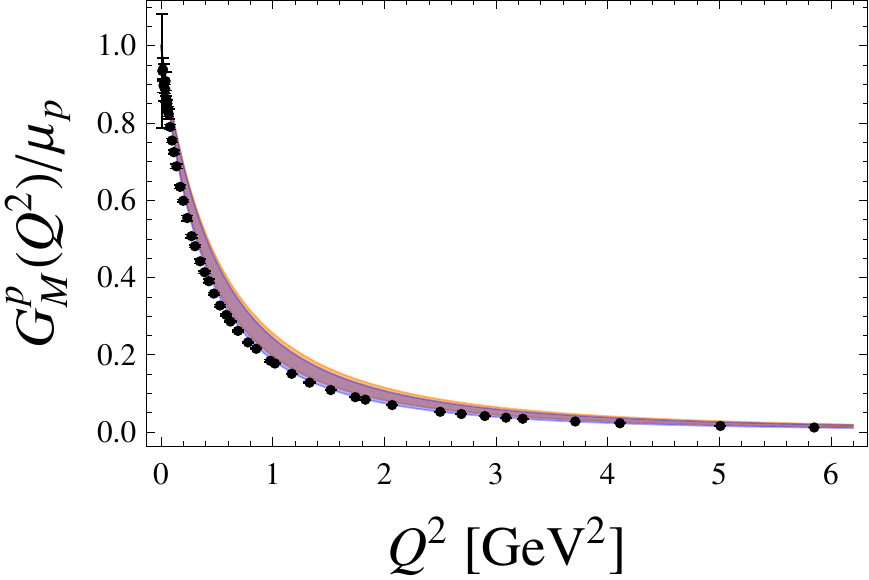}
\includegraphics[width=7cm]{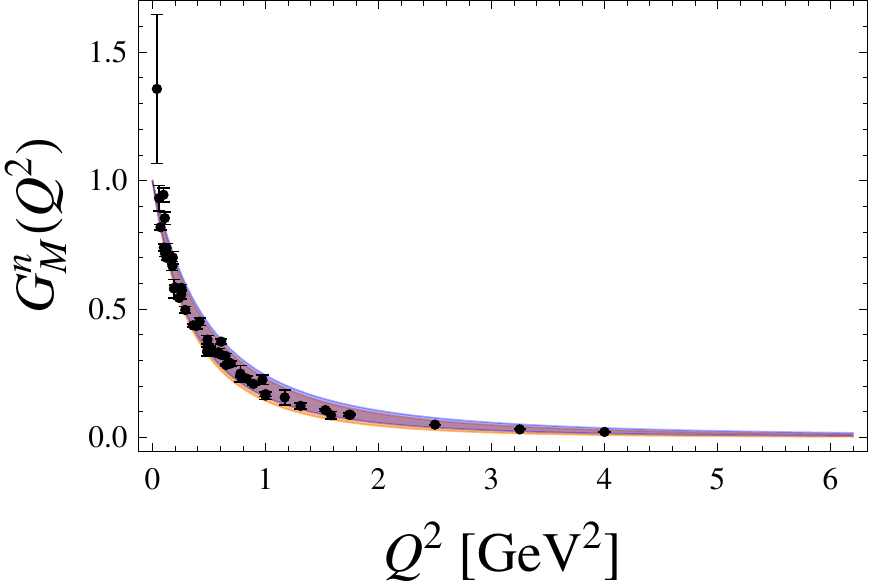}
\caption{(color online). The electromagnetic nucleon form factors compared to the data for 
the proton \cite{Arrington:2007ux} and neutron \cite{Perdrisat:2006hj} (and references therein). Orange band: $g_{\omega NN}= 9 $, blue band: $g_{\omega NN}= 12$. }
\label{fig:femN}
\end{figure*}

The result of varying the masses according to the half-width rule is
presented in Fig.~\ref{fig:femN}. We consider the $SU(3)$, case where
$g_{\omega NN} /g_{\rho NN}= f_{\omega NN} /f_{\rho NN}= 3$, as well
as a $30\%$ violation for the ratio (orange and blue band in
Fig.~\ref{fig:femN}, respectively). We recall that in the
meson-exchange models~\cite{Machleidt:1987hj} or in dispersive
analyses of the nucleon form
factors~\cite{Mergell:1995bf,Belushkin:2006qa} even much
larger symmetry breaking is needed to comply with the phenomenology.

\subsection{Axial and pseudoscalar form factors}

\begin{figure*}[tbc]
\includegraphics[width=7cm]{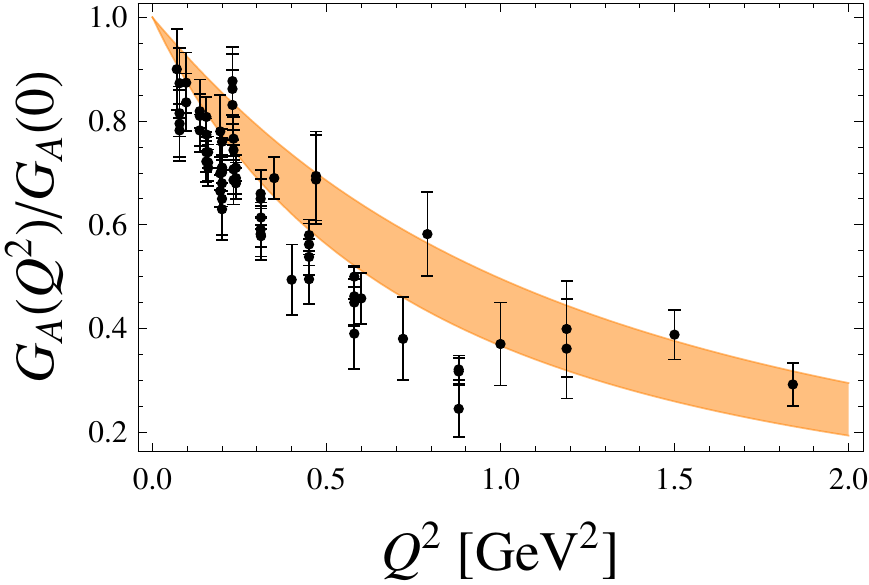}
\includegraphics[width=7cm]{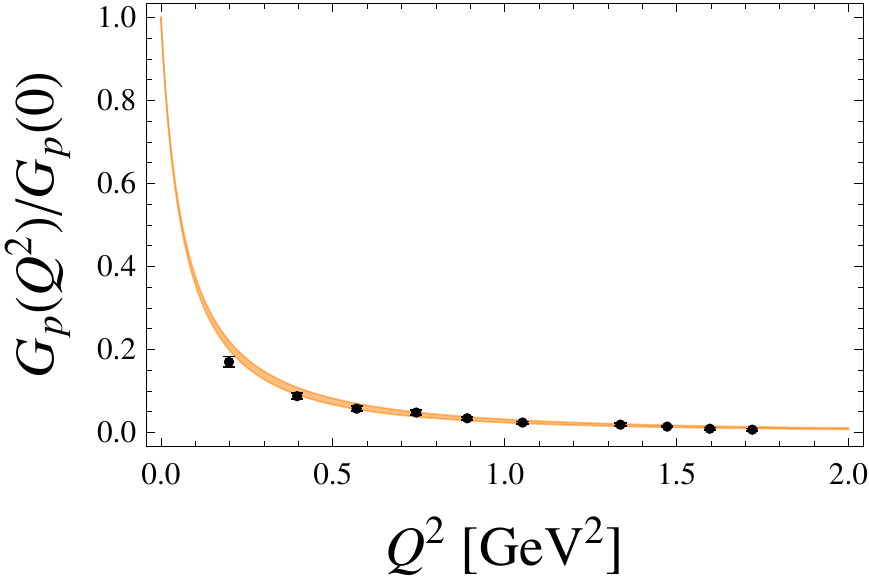}
\caption{(color online). The isovector-axial nucleon form factors using two
  axial masses (left panel). The isovector-pseudoscalar nucleon form factor
  using two axial masses and the lightest pion considered in \cite{Alexandrou:2010hf} (right panel).
\label{fig:fweakN}}
\end{figure*}

The axial matrix element of nucleon is defined by (for a recent review
see, e.g., Ref.~\cite{Bernard:2001rs} and references therein)
\begin{eqnarray}
&&\langle N(p') | J_A^{\mu a} (0) | N(p) \rangle \nonumber = \\
&& \bar u (p')
 \frac{ \tau^a}2 \gamma_5 \left[ \gamma_\mu G_A (q^2) + \frac{q^\mu}{2M_N}
 G_P(q^2) \right] u (p),
\end{eqnarray} 
where $G_A(q^2)$ and $G_P(q^2) $ are the axial and the induced
pseudoscalar form factors, respectively.  The QCD axial current is
\begin{eqnarray}
J_A^{\mu a} (x)= \bar q (x) \gamma^\mu \frac{\tau^a}{2} q (x),   
\end{eqnarray} 
while current-field identity relating it to the axial-meson field $A^{\mu \nu}$ and the 
pseudoscalar field $P$ is 
\begin{eqnarray}
J_A^{\mu a} (x) = \sum f_A \partial_\nu A^{a \mu \nu}(x) + \sum f_P
\partial^\mu P^a(x). 
\end{eqnarray}
Therefore we have
\begin{eqnarray}
G_A (t) &=& g_A + \sum_A \frac{f_A g_{ANN} t}{M_A^2-t}, \nonumber \\  
G_P (t) &=& -\sum_A \frac{4 M_N^2 f_A g_{ANN}}{M_A^2-t}
+   \sum_P \frac{4 M_N F_P g_{PNN}}{M_P^2-t}. \nonumber \\
\end{eqnarray}
We use here the extended PCAC form~\cite{Dominguez:1976ut}, which for 
the on-shell mesons reads 
\begin{eqnarray}
\partial_\mu J_A^{\mu a} (x) =  \sum f_P M_P^2 P^a(x), 
\end{eqnarray} 
yielding the following relation among the form factors:  
\begin{eqnarray}
2 M_N G_A (t) + \frac{t}{2 M_N} G_P (t) = \sum_P \frac{2 M_P^2
F_P}{M_P^2-t} g_{PNN}.
\end{eqnarray} 
The pseudoscalar-nucleon coupling is defined by 
\begin{eqnarray}
\langle p' | (\partial^2 +M_P^2 ) P_n^a (x) | p \rangle = g_{PNN} \bar u(p') i \gamma_5 \tau^a u(p).
\end{eqnarray}
>From here we get the (extended) Goldberger-Treiman relation
\begin{eqnarray}
M_N g_A = \sum_P F_P g_{PNN} = f_\pi g_{\pi NN} + f_{\pi'} g_{\pi'
NN}+ \dots . \nonumber \\
\end{eqnarray} 

The high-energy behavior of the weak form factors in QCD was discussed
many years ago~\cite{Alvegard:1979ui,Brodsky:1980sx}.  At high $Q^2$, one
has for the isovector~\cite{Carlson:1985zu} and
isoscalar~\cite{Carlson:1987en} the asymptotic behavior,
\begin{eqnarray}
Q^4 G_A(-Q^2) \to {\rm const}.
\end{eqnarray}
We also have the sum rules
\begin{eqnarray}
g_A &=& \sum_A f_A g_{ANN}, \nonumber \\
0 &=& \sum_A f_A g_{ANN} M_A^2. 
\end{eqnarray}

%\begin{eqnarray}
%J_A^{\mu a} (x) = \sum \frac{M_A^2}{f_A} a^\mu +  \sum f_P \partial^\mu P 
%\end{eqnarray} 

It is noteworthy that most  determinations of the axial form factor 
proceed via a dipole fit, 
\begin{eqnarray}
G_A(t) = \frac{g_A}{(1-t/\Lambda_A^2)^2},
\label{eq:ga_dip}
\end{eqnarray} 
suggesting a $1/t^2$ fall off at large $t$. 
The values of the parameter are
$\Lambda_A= 1.026(21)$~GeV or $\Lambda_A= 1.069(16)$~GeV, depending on
the process~\cite{Bernard:2001rs}. In the literature $\Lambda_A$ is
denoted and called the {\em axial mass} (see, e.g., ~\cite{Bernard:2001rs}). This
is not our axial meson mass, since Eq.~(\ref{eq:ga_dip}), although
phenomenologically successful, cannot be justified from a field-%
theoretic point of view and is in contradiction with the large-$N_c$
motivated parameterization.

The minimum meson-dominance ansatz compatible with low- and high-energy
constraints reads 
\begin{eqnarray}\label{GaGpeq}
G_A(t) &=& g_A \frac{m_{a_1}^2 m_{a_1'}^2}{(m_{a_1}^2-t)(m_{a_1'}^2-t)}, \nonumber \\ 
G_P(t) &=& G_A(t) \frac{G_P(0)}{m_p^2-t}.
%G_P(t) &=& g_P \frac{m_\pi^2 m_{\pi'}^2 m_{\pi''}^2 }
%{(m_\pi^2-t)( m_{\pi'}^2-t)(m_{\pi''}^2-t)}
\end{eqnarray} 
By applying the half-width rule to this parameterization, i.e., using $m_{a_1}=1.230$~GeV, $m_{a_1'}=1.647$, $\Gamma_{a_1}=0.425$~GeV, and $\Gamma_{a_1'}=0.254$~GeV, we get the
results depicted in Fig.~\ref{fig:fweakN}. As we can see, the results are
in reasonable agreement with the data. Actually, the two axial mesons
are incorporated as a product of monopoles, but since they have an
overlapping spectrum, the net effect is essentially a dipole form factor
with an average mass which is somewhat larger than the usual dipole cut-off.

\subsection{Gravitational form factors}

The discussion of the nucleon gravitational form factors follows closely
the pion case with suitable changes. For the nucleon case, the quark
contributions to these form factors have been determined by the QCDSF
Collaboration~~\cite{Gockeler:2003jfa} and the LHPC
Collaboration~\cite{Hagler:2007hu}.

The decomposition, corresponding to the energy-momentum tensor matrix
elements taken between nucleon states, reads
\begin{eqnarray}
\langle p^\prime| \Theta_{\mu\nu}^{q} |p\rangle &=& \bar u(p^\prime)\biggl[
    A^{q}_{20}(t)\,\frac{\gamma_\mu P_\nu+\gamma_\nu P_\mu}{2} \nonumber \\ 
&+&  B^{q}_{20}(t)\,\frac{i(P_{\mu}\sigma_{\nu\rho}+P_{\nu}\sigma_{\mu\rho})
    \Delta^\rho}{4M_N}  \nonumber \\
    &+& C^{q}_{20}(t)\,\frac{\Delta_\mu\Delta_\nu-g_{\mu\nu}\Delta^2}{M_N}
    \biggr]u(p),  \label{emt:N} 
\end{eqnarray}
where $\sigma^{\mu\nu}= \frac{i}{2} [ \gamma^\mu, \gamma^\nu] $
(the Bjorken-Drell notation), the scalar functions are moments of the GPDs,
the momentum transfer is denoted as $\Delta=p'-p$, and 
the average nucleon momentum is $P= (p'+p)/2$. Taking the trace and applying the
Gordon identity, $2 M_N \bar u(p') \gamma^\mu u(p) = \bar u(p') ( i
\sigma^{\mu\rho} \Delta_\rho + 2P^\mu ) u(p) $, as well as the Dirac equation,
$(\slashchar{p}-M_N)u(p)=0$ and $\bar u(p')(\slashchar{p'}-M_N)=0$, we obtain
the following expression for the spin-0 gravitational form factor of the nucleon:
\begin{eqnarray}
\!\!\!\Theta_N^q( t) = M_N \left[ A_{20}^q (t) + \frac{t}{4M_M^2} B_{20}^q (t) -
\frac{3t}{M_N^2} C_{20}^q (t) \right] , \nonumber \\
\label{eq:thetaNq}
\end{eqnarray} 
whereas the spin-2 (normalized) component becomes 
\begin{eqnarray}
\!\!\!F_T^q( t) = \frac{A_{20}^q (t)}{A_{20}^q (0)}
\label{eq:FT}
\end{eqnarray}

\begin{figure}[htbp]
\begin{center}
\includegraphics[width=7cm]{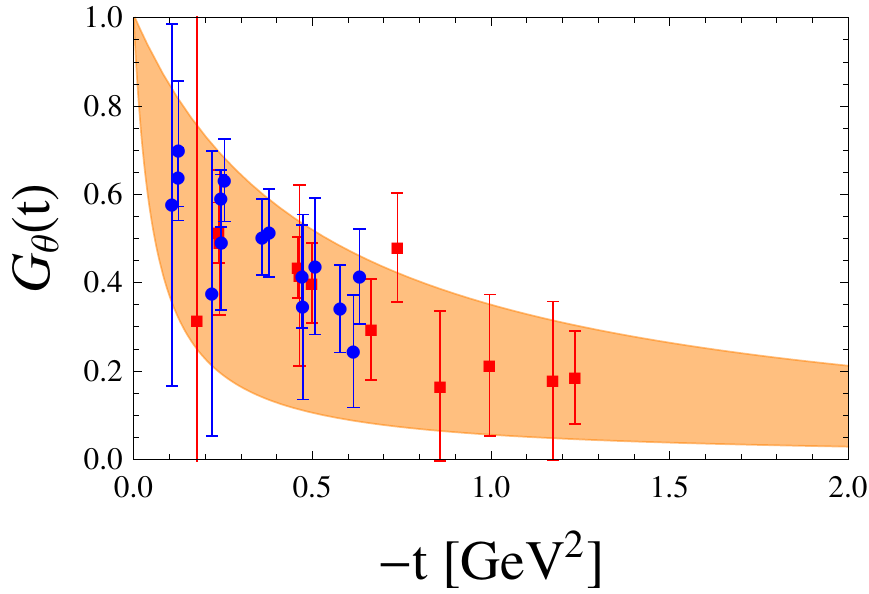}
\caption{(color online). Spin-0 gravitational form factor of the nucleon, $G_{\theta}(t)$ obtained from the lattice simulations of Ref.\cite{Hagler:2007hu} at the pion masses $m_{\pi}=352.3$MeV (blue circles) and $m_{\pi}=356.6$~MeV (red squares). }
\label{GFFnucleon}
\end{center}
\end{figure}

\begin{figure}[htbp]
\begin{center}
\includegraphics[width=7cm]{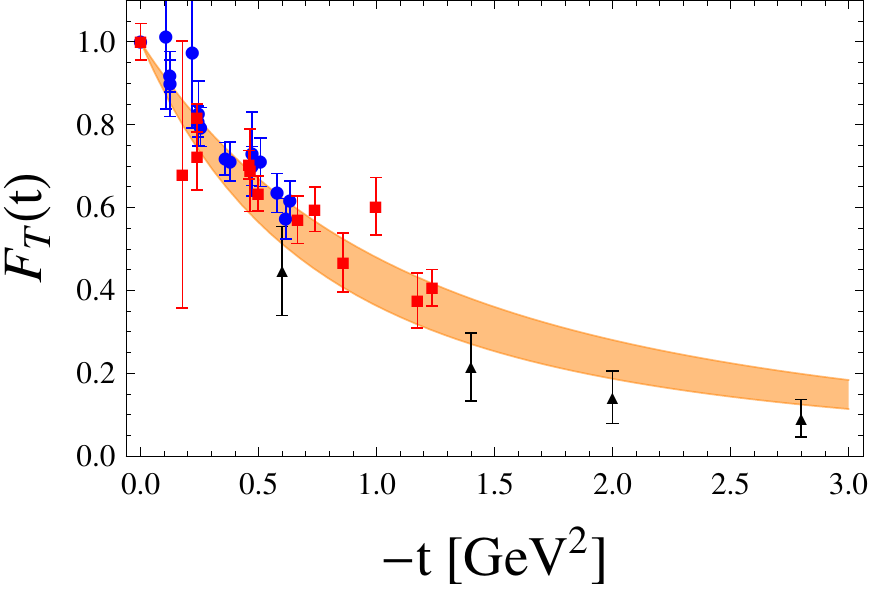}
\caption{(color online). Spin-2 isoscalar gravitational form factor of the nucleon,
  $F_T (t)$ obtained from the lattice simulations of
  Ref.~\cite{Hagler:2007hu} at the pion masses $m_{\pi}=352.3$~MeV  and $m_{\pi}=356.6$~MeV (blue circles and red squares respectively), together with the 
results of Ref.~\cite{Hagler:2007hu} linearly extrapolated to the physical pion mass (black triangles).}
\label{GFFnucleon2}
\end{center}
\end{figure}
The trace is dominated by scalar,$0^{++}$, states. For the monopole
representation of the gravitational form factor, 
\begin{eqnarray}
 G_ {\theta}(t) = \frac{m_{f_0}^2}{m_{f_0}^2-t} ,
\end{eqnarray} 
depicted in Fig.~\ref{GFFnucleon}, we use the updated values of the sigma
meson  properties from the latest edition of the PDG Tables~\cite{PDG2012}, i.e.,
$m_{\sigma}=475$~MeV and $\Gamma_{\sigma}=550$~MeV.

The traceless part of the EM tensor corresponds to a spin-2 isoscalar
gravitational form factor which naturally couples to the $f_2$ meson
within a tensor dominance approximation. For the nucleon this FF has
been determined by two lattice groups in
Refs.~\cite{Gockeler:2003jfa,Hagler:2007hu}. Actually, in
Ref.~\cite{Gockeler:2003jfa} a dipole fit describes the data successfully, namely
\begin{eqnarray}
 F_T(t) = \frac{1}{(1-t/\Lambda_{T}^2)^2} ,
\end{eqnarray} 
with $\Lambda_T=1.1(2)$~GeV, if a linear extrapolation in $m_\pi$ to the
physical point is assumed. Assuming an asymptotic falloff for the form
factor, such that $F_T(t) = {\cal O} (t^{-2})$, we just take a sum of two
monopoles that reduces to
\begin{eqnarray}
 F_T(t) = \frac{m_{f_2}^2}{m_{f_2}^2-t} \frac{m_{f_2'}^2}{m_{f_2'}^2-t} .
\end{eqnarray} 
The PDG Tables quote $m_{f_2}=1.320$~GeV and $\Gamma_{f_2}=0.185$~GeV, and for
the first excited state gives $m_{f_2'}=1.525$~GeV and
$\Gamma_{f_2'}=0.073$~GeV.~\footnote{This is somewhat different from
  the results of Ref.~\cite{Anisovich:2011in} with about four times
  larger width for the $f_2(1565)$ meson. In any case, for the spin-2
  form factor the difference becomes rather irrelevant in the region
  below 1~GeV.}

As we can see from Fig.~\ref{GFFnucleon2}, similarly to the case of
the axial nucleon FF, the dipole FF with an uncertainty essentially
corresponds to two monopoles after the half-width rule has been
implemented.

\section{Conclusions}
\label{sec:concl}

In the present work we have taken advantage of the well-known fact
that in the large-$N_c$ limit of QCD the generalized hadronic form factors,
probing bilinear $\bar q q $ operators with given $J^{PC}$ quantum
numbers, feature generalized meson dominance of $\bar q q $ states
with the same quantum numbers. They assume the monopole form,
\begin{eqnarray}
\langle A(p') | J (0) | \langle B(p) \rangle \sim 
\sum_{n} \frac{ c_n^{AB} m_n^2}{m_n^2-t},
\end{eqnarray}
where $m_n$ are the meson masses and $c_n^{AB}$ suitable couplings. 
Thus generalized form factors at some finite momentum transfer essentially
measure the masses of the lowest lying mesons. 

The goal of this paper was to present a comprehensive analysis of the
pion and nucleon form factors, providing a theoretical uncertainty
following from the half-width rule.  We have incorporated
\begin{itemize}
\item The correct asymptotic power like behavior of form factors
  (short-distance constraints).
\item Low-momentum normalization constraints.
\item Minimum number of mesons with the relevant quantum numbers in
  each channel.
\item Theoretical error estimates based the half-width rule. 
\end{itemize}
Given the approximate nature of the underlying large-$N_c$ expansion,
we should not expect perfect agreement with data. Rather, the
addressed question is how we can estimate the accuracy of the
large-$N_c$ expansion for hadronic form factors, which are basic
experimental quantities. In the present paper we provide arguments in
favor of the simple rule, where a rough estimate is given by varying the
resonance mass within its width range. As we have seen, this provides
a surprisingly close answer to the data, which fall within the bands
produced with the half-width rule.  While this presumably is a
conservative assumption, it still provides a cheap estimate based on an
independent source of information.

Extension to other baryons and mesons is straightforward, as well as to
the transition matrix elements. One useful application of the meson-%
dominance scheme is the {\it a priori} determination of Generalized
Form Factors, just based on the PDG Tables, for which abundant data start to be
produced on the lattice. Our calculations show that improving on the
uncertainty predictions based of the half-width rule may require
highly refined lattice studies beyond the present accuracy.

\appendix

\begin{figure}[tbc]
\includegraphics[width=7cm]{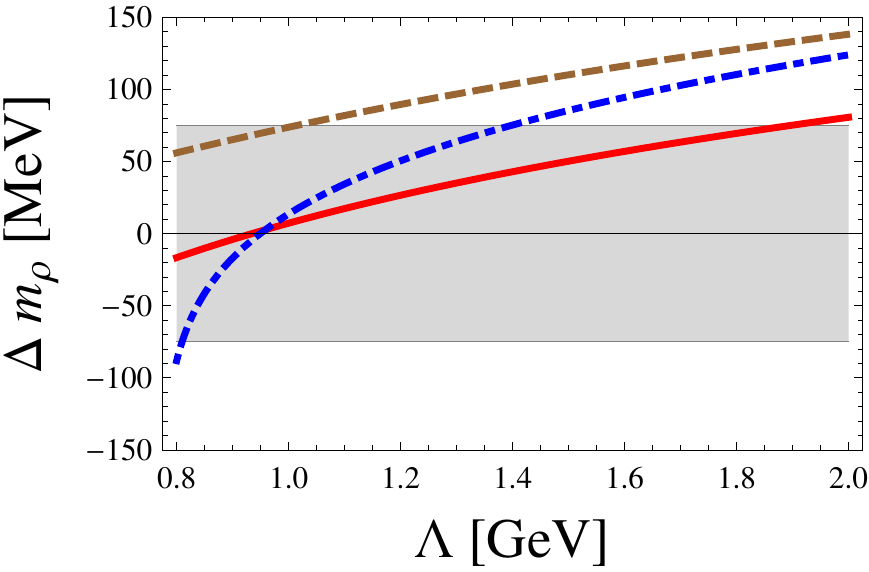}
\includegraphics[width=7cm]{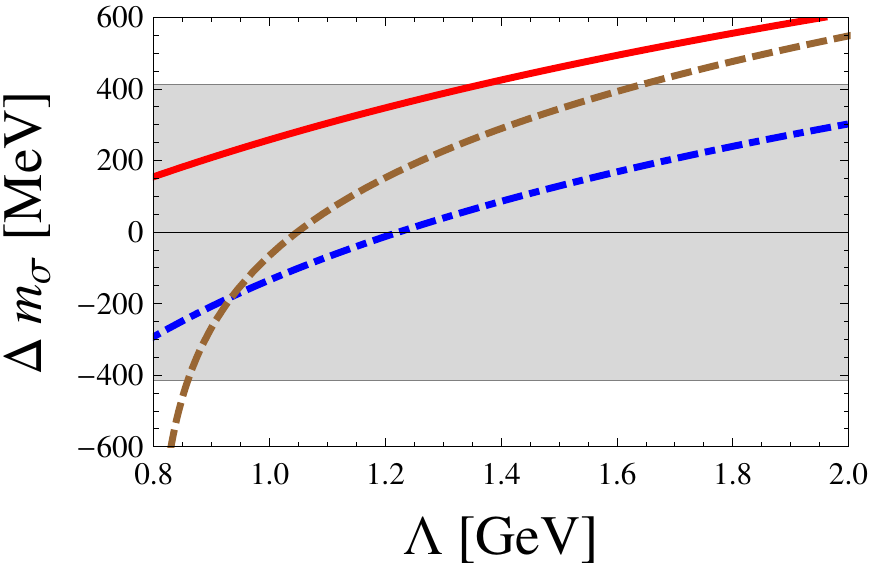}
\caption{(color online). Mass-shifts $\Delta m$ (in GeV) of the $\rho$-meson (top
  panel) and the $\sigma$-meson (bottom panel), due to pion loops, plotted as
  functions of the cut-off $\Lambda$ (in GeV) for several cut-off
  functions: monopole (dashed brown), dipole (solid red), sharp cut-off (dot-dashed blue).}
\label{fig:dm}
\end{figure}

\section{Cut-off dependence of the  mass shift}
\label{sec:cut-off}
In this appendix we analyze the mass shift of the lowest lying scalar
and vector resonances due to the most relevant decays $S \to \pi \pi$
and $V \to \pi \pi$.  With a chiral derivative coupling, one
obtains~\cite{Nieves:2011gb}
\begin{eqnarray}
\Gamma_S (s) &=& \frac{3 m_S s}{16 \pi f_\pi^4} \rho_S \left[ c_d + (c_m -c_d)
\frac{2m_\pi^2}{m_S^2}\right]^2, \nonumber \\ 
\Gamma_V (s) &=& \frac{G_V^2 m_V s}{48\pi f_\pi^4} \rho_V^3,
\end{eqnarray} 
where $\rho_R =\rho(m_R^2)= \sqrt{1-4m_\pi^2/m_R^2}$. Note the
additional $s$ factors appearing in the widths. By using the
dispersion relation for the self-energy, Eq.~(\ref{eq:self-disp}), we
can see that the real part of the integral is quartically divergent
and thus three subtractions are needed. This is equivalent to fixing
the mass and the width. On the contrary, the imaginary part is finite
and provides the width at the physical value. This argument shows that
we need more input information and until then have no predictive
power. Within a large $N_c$ environment~\cite{Nieves:2011gb} we use BW
and not pole reference subtraction values for both mesons in this
study.

In order to get an estimate, we assume some form of the hadronic form factor for
the vertex $\rho \pi \pi$.  Out of ignorance, we consider different
cut-off functions
\begin{eqnarray}
G_{\rho\pi\pi}(q^2)=\left(\frac{q^2+\Lambda^2}{s+\Lambda^2} \right)^n,
\end{eqnarray}
which preserve the imaginary part, $G_{\rho \pi\pi} (m_\rho^2)$, and
hence the width.  Here $n=1$ corresponds to a monopole, $n=2$
corresponds to a dipole, and the limit $n \to \infty$ corresponds to a
sharp cut-off. We naturally expect $\Lambda$ in the range above
$m_\rho$ and around $m_{\rho'}$, which corresponds to ignoring all
states above these energies.
With these conditions, and assuming a small correction, we get a mass shift 
\begin{eqnarray}
\Delta m_\rho = \frac1{\pi} \int_{4 m_\pi^2}^\infty ds  \frac{G_{\rho \pi\pi}(s) \Gamma_{\rho \pi \pi}(s)}{s(s-m_\rho^2)}.
\end{eqnarray}
The results are depicted in Fig.~\ref{fig:dm} as a function of the
cut-off scale. As expected, the overall order of magnitude is quite
compatible with the half-width rule.

In a theory with a finite number of resonances there is, of course, an
implicit high energy cut-off corresponding to the next (not included)
meson. Assuming a vanishing contribution of the higher energy states,
one may make a rough estimate of the mass shift. For instance, for the
$\rho$-meson one has ~\cite{Leinweber:1993yw} a shift of about half
the width.  Of course, this may be partly a numerical coincidence, but
it illustrates the point that parametrically the mass-shift and the
width scale in a similar way.


\begin{thebibliography}{98}%
\makeatletter
\providecommand \@ifxundefined [1]{%
 \@ifx{#1\undefined}
}%
\providecommand \@ifnum [1]{%
 \ifnum #1\expandafter \@firstoftwo
 \else \expandafter \@secondoftwo
 \fi
}%
\providecommand \@ifx [1]{%
 \ifx #1\expandafter \@firstoftwo
 \else \expandafter \@secondoftwo
 \fi
}%
\providecommand \natexlab [1]{#1}%
\providecommand \enquote  [1]{``#1''}%
\providecommand \bibnamefont  [1]{#1}%
\providecommand \bibfnamefont [1]{#1}%
\providecommand \citenamefont [1]{#1}%
\providecommand \href@noop [0]{\@secondoftwo}%
\providecommand \href [0]{\begingroup \@sanitize@url \@href}%
\providecommand \@href[1]{\@@startlink{#1}\@@href}%
\providecommand \@@href[1]{\endgroup#1\@@endlink}%
\providecommand \@sanitize@url [0]{\catcode `\\12\catcode `\$12\catcode
  `\&12\catcode `\#12\catcode `\^12\catcode `\_12\catcode `\%12\relax}%
\providecommand \@@startlink[1]{}%
\providecommand \@@endlink[0]{}%
\providecommand \url  [0]{\begingroup\@sanitize@url \@url }%
\providecommand \@url [1]{\endgroup\@href {#1}{\urlprefix }}%
\providecommand \urlprefix  [0]{URL }%
\providecommand \Eprint [0]{\href }%
\providecommand \doibase [0]{http://dx.doi.org/}%
\providecommand \selectlanguage [0]{\@gobble}%
\providecommand \bibinfo  [0]{\@secondoftwo}%
\providecommand \bibfield  [0]{\@secondoftwo}%
\providecommand \translation [1]{[#1]}%
\providecommand \BibitemOpen [0]{}%
\providecommand \bibitemStop [0]{}%
\providecommand \bibitemNoStop [0]{.\EOS\space}%
\providecommand \EOS [0]{\spacefactor3000\relax}%
\providecommand \BibitemShut  [1]{\csname bibitem#1\endcsname}%
\let\auto@bib@innerbib\@empty
%</preamble>
\bibitem [{\citenamefont {Miller}(2010)}]{Miller:2010nz}%
  \BibitemOpen
  \bibfield  {author} {\bibinfo {author} {\bibfnamefont {G.~A.}\ \bibnamefont
  {Miller}},\ }\href {\doibase
  http://www.annualreviews.org/doi/pdf/10.1146/annurev.nucl.012809.104508}
  {\bibfield  {journal} {\bibinfo  {journal} {Ann. Rev. Nucl. Part. Sci.}\
  }\textbf {\bibinfo {volume} {60}},\ \bibinfo {pages} {1} (\bibinfo {year}
  {2010})},\ \Eprint {http://arxiv.org/abs/1002.0355} {arXiv:1002.0355
  [nucl-th]} \BibitemShut {NoStop}%
%%CITATION = 1002.0355;%%
\bibitem [{\citenamefont {Drechsel}\ and\ \citenamefont
  {Walcher}(2008)}]{Drechsel:2007sq}%
  \BibitemOpen
  \bibfield  {author} {\bibinfo {author} {\bibfnamefont {D.}~\bibnamefont
  {Drechsel}}\ and\ \bibinfo {author} {\bibfnamefont {T.}~\bibnamefont
  {Walcher}},\ }\href {\doibase 10.1103/RevModPhys.80.731} {\bibfield
  {journal} {\bibinfo  {journal} {Rev. Mod. Phys.}\ }\textbf {\bibinfo {volume}
  {80}},\ \bibinfo {pages} {731} (\bibinfo {year} {2008})},\ \Eprint
  {http://arxiv.org/abs/0711.3396} {arXiv:0711.3396 [hep-ph]} \BibitemShut
  {NoStop}%
%%CITATION = 0711.3396;%%
\bibitem [{\citenamefont {Brodsky}\ and\ \citenamefont
  {Chertok}(1976)}]{Brodsky:1976rz}%
  \BibitemOpen
  \bibfield  {author} {\bibinfo {author} {\bibfnamefont {S.~J.}\ \bibnamefont
  {Brodsky}}\ and\ \bibinfo {author} {\bibfnamefont {B.}~\bibnamefont
  {Chertok}},\ }\href {\doibase 10.1103/PhysRevD.14.3003} {\bibfield  {journal}
  {\bibinfo  {journal} {Phys.Rev.}\ }\textbf {\bibinfo {volume} {D14}},\
  \bibinfo {pages} {3003} (\bibinfo {year} {1976})}\BibitemShut {NoStop}%
\bibitem [{\citenamefont {Lepage}\ and\ \citenamefont
  {Brodsky}(1980)}]{Lepage:1980fj}%
  \BibitemOpen
  \bibfield  {author} {\bibinfo {author} {\bibfnamefont {G.~P.}\ \bibnamefont
  {Lepage}}\ and\ \bibinfo {author} {\bibfnamefont {S.~J.}\ \bibnamefont
  {Brodsky}},\ }\href {\doibase 10.1103/PhysRevD.22.2157} {\bibfield  {journal}
  {\bibinfo  {journal} {Phys. Rev.}\ }\textbf {\bibinfo {volume} {D22}},\
  \bibinfo {pages} {2157} (\bibinfo {year} {1980})}\BibitemShut {NoStop}%
%%CITATION = PHRVA,D22,2157;%%
\bibitem [{\citenamefont {Radyushkin}(2000)}]{Radyushkin:2000uy}%
  \BibitemOpen
  \bibfield  {author} {\bibinfo {author} {\bibfnamefont {A.~V.}\ \bibnamefont
  {Radyushkin}},\ }\href@noop {} {\  (\bibinfo {year} {2000})},\ \Eprint
  {http://arxiv.org/abs/hep-ph/0101225} {hep-ph/0101225} \BibitemShut {NoStop}%
%%CITATION = HEP-PH/0101225;%%
\bibitem [{\citenamefont {Goeke}\ \emph {et~al.}(2001)\citenamefont {Goeke},
  \citenamefont {Polyakov},\ and\ \citenamefont
  {Vanderhaeghen}}]{Goeke:2001tz}%
  \BibitemOpen
  \bibfield  {author} {\bibinfo {author} {\bibfnamefont {K.}~\bibnamefont
  {Goeke}}, \bibinfo {author} {\bibfnamefont {M.~V.}\ \bibnamefont {Polyakov}},
  \ and\ \bibinfo {author} {\bibfnamefont {M.}~\bibnamefont {Vanderhaeghen}},\
  }\href@noop {} {\bibfield  {journal} {\bibinfo  {journal} {Prog. Part. Nucl.
  Phys.}\ }\textbf {\bibinfo {volume} {47}},\ \bibinfo {pages} {401} (\bibinfo
  {year} {2001})},\ \Eprint {http://arxiv.org/abs/hep-ph/0106012}
  {hep-ph/0106012} \BibitemShut {NoStop}%
%%CITATION = HEP-PH/0106012;%%
\bibitem [{\citenamefont {Ji}(2004)}]{Ji:2004gf}%
  \BibitemOpen
  \bibfield  {author} {\bibinfo {author} {\bibfnamefont {X.-D.}\ \bibnamefont
  {Ji}},\ }\href@noop {} {\bibfield  {journal} {\bibinfo  {journal} {Ann. Rev.
  Nucl. Part. Sci.}\ }\textbf {\bibinfo {volume} {54}},\ \bibinfo {pages} {413}
  (\bibinfo {year} {2004})}\BibitemShut {NoStop}%
%%CITATION = ARNUA,54,413;%%
\bibitem [{\citenamefont {Hagler}(2010)}]{Hagler:2009ni}%
  \BibitemOpen
  \bibfield  {author} {\bibinfo {author} {\bibfnamefont {P.}~\bibnamefont
  {Hagler}},\ }\href@noop {} {\bibfield  {journal} {\bibinfo  {journal}
  {Phys.Rept.}\ }\textbf {\bibinfo {volume} {490}},\ \bibinfo {pages} {49}
  (\bibinfo {year} {2010})},\ \Eprint {http://arxiv.org/abs/0912.5483}
  {arXiv:0912.5483 [hep-lat]} \BibitemShut {NoStop}%
%%CITATION = ARXIV:0912.5483;%%
\bibitem [{\citenamefont {Sakurai}(1969)}]{sakurai1969currents}%
  \BibitemOpen
  \bibfield  {author} {\bibinfo {author} {\bibfnamefont {J.}~\bibnamefont
  {Sakurai}},\ }\href@noop {} {\emph {\bibinfo {title} {Currents and mesons}}}\
  (\bibinfo  {publisher} {University of Chicago Press},\ \bibinfo {year}
  {1969})\BibitemShut {NoStop}%
\bibitem [{\citenamefont {O'Connell}\ \emph {et~al.}(1997)\citenamefont
  {O'Connell}, \citenamefont {Pearce}, \citenamefont {Thomas},\ and\
  \citenamefont {Williams}}]{O'Connell:1995wf}%
  \BibitemOpen
  \bibfield  {author} {\bibinfo {author} {\bibfnamefont {H.~B.}\ \bibnamefont
  {O'Connell}}, \bibinfo {author} {\bibfnamefont {B.~C.}\ \bibnamefont
  {Pearce}}, \bibinfo {author} {\bibfnamefont {A.~W.}\ \bibnamefont {Thomas}},
  \ and\ \bibinfo {author} {\bibfnamefont {A.~G.}\ \bibnamefont {Williams}},\
  }\href {\doibase 10.1016/S0146-6410(97)00044-6} {\bibfield  {journal}
  {\bibinfo  {journal} {Prog. Part. Nucl. Phys.}\ }\textbf {\bibinfo {volume}
  {39}},\ \bibinfo {pages} {201} (\bibinfo {year} {1997})},\ \Eprint
  {http://arxiv.org/abs/hep-ph/9501251} {arXiv:hep-ph/9501251} \BibitemShut
  {NoStop}%
%%CITATION = HEP-PH/9501251;%%
\bibitem [{\citenamefont {Masjuan}\ \emph {et~al.}(2008)\citenamefont
  {Masjuan}, \citenamefont {Peris},\ and\ \citenamefont
  {Sanz-Cillero}}]{Masjuan:2008fv}%
  \BibitemOpen
  \bibfield  {author} {\bibinfo {author} {\bibfnamefont {P.}~\bibnamefont
  {Masjuan}}, \bibinfo {author} {\bibfnamefont {S.}~\bibnamefont {Peris}}, \
  and\ \bibinfo {author} {\bibfnamefont {J.}~\bibnamefont {Sanz-Cillero}},\
  }\href {\doibase 10.1103/PhysRevD.78.074028} {\bibfield  {journal} {\bibinfo
  {journal} {Phys.Rev.}\ }\textbf {\bibinfo {volume} {D78}},\ \bibinfo {pages}
  {074028} (\bibinfo {year} {2008})},\ \Eprint {http://arxiv.org/abs/0807.4893}
  {arXiv:0807.4893 [hep-ph]} \BibitemShut {NoStop}%
%%CITATION = ARXIV:0807.4893;%%
\bibitem [{\citenamefont {Masjuan}(2009)}]{Masjuan:2008xg}%
  \BibitemOpen
  \bibfield  {author} {\bibinfo {author} {\bibfnamefont {P.}~\bibnamefont
  {Masjuan}},\ }\href {\doibase 10.1016/j.nuclphysbps.2008.12.035} {\bibfield
  {journal} {\bibinfo  {journal} {Nucl.Phys.Proc.Suppl.}\ }\textbf {\bibinfo
  {volume} {186}},\ \bibinfo {pages} {149} (\bibinfo {year} {2009})},\ \Eprint
  {http://arxiv.org/abs/0809.2704} {arXiv:0809.2704 [hep-ph]} \BibitemShut
  {NoStop}%
%%CITATION = ARXIV:0809.2704;%%
\bibitem [{\citenamefont {Masjuan}(2012)}]{Masjuan:2012wy}%
  \BibitemOpen
  \bibfield  {author} {\bibinfo {author} {\bibfnamefont {P.}~\bibnamefont
  {Masjuan}},\ }\href@noop {} {\  (\bibinfo {year} {2012})},\ \Eprint
  {http://arxiv.org/abs/1206.2549} {arXiv:1206.2549 [hep-ph]} \BibitemShut
  {NoStop}%
%%CITATION = ARXIV:1206.2549;%%
\bibitem [{\citenamefont {Ruiz~Arriola}\ and\ \citenamefont
  {Broniowski}(2010)}]{RuizArriola:2010fj}%
  \BibitemOpen
  \bibfield  {author} {\bibinfo {author} {\bibfnamefont {E.}~\bibnamefont
  {Ruiz~Arriola}}\ and\ \bibinfo {author} {\bibfnamefont {W.}~\bibnamefont
  {Broniowski}},\ }\href {\doibase 10.1103/PhysRevD.81.054009} {\bibfield
  {journal} {\bibinfo  {journal} {Phys.Rev.}\ }\textbf {\bibinfo {volume}
  {D81}},\ \bibinfo {pages} {054009} (\bibinfo {year} {2010})},\ \Eprint
  {http://arxiv.org/abs/1001.1636} {arXiv:1001.1636 [hep-ph]} \BibitemShut
  {NoStop}%
%%CITATION = ARXIV:1001.1636;%%
\bibitem [{\citenamefont {Arriola}\ and\ \citenamefont
  {Broniowski}(2011{\natexlab{a}})}]{Arriola:2010aj}%
  \BibitemOpen
  \bibfield  {author} {\bibinfo {author} {\bibfnamefont {E.~R.}\ \bibnamefont
  {Arriola}}\ and\ \bibinfo {author} {\bibfnamefont {W.}~\bibnamefont
  {Broniowski}},\ }\href {\doibase 10.1063/1.3575030} {\bibfield  {journal}
  {\bibinfo  {journal} {AIP Conf.Proc.}\ }\textbf {\bibinfo {volume} {1343}},\
  \bibinfo {pages} {361} (\bibinfo {year} {2011}{\natexlab{a}})},\ \Eprint
  {http://arxiv.org/abs/1011.5176} {arXiv:1011.5176 [hep-ph]} \BibitemShut
  {NoStop}%
%%CITATION = ARXIV:1011.5176;%%
\bibitem [{\citenamefont {Arriola}\ and\ \citenamefont
  {Broniowski}(2011{\natexlab{b}})}]{Arriola:2011en}%
  \BibitemOpen
  \bibfield  {author} {\bibinfo {author} {\bibfnamefont {E.~R.}\ \bibnamefont
  {Arriola}}\ and\ \bibinfo {author} {\bibfnamefont {W.}~\bibnamefont
  {Broniowski}},\ }\href@noop {} {\bibfield {journal}{\bibinfo{journal}{Bled Workshops in Physics. Vol. 12 No. 1}} ,\ \bibinfo {pages} {7} (\bibinfo {year}
  {2011}{\natexlab{b}})},\ \Eprint {http://arxiv.org/abs/1110.2863}
  {arXiv:1110.2863 [hep-ph]} \BibitemShut {NoStop}%
%%CITATION = ARXIV:1110.2863;%%
\bibitem [{\citenamefont {Masjuan}\ \emph
  {et~al.}(2012{\natexlab{a}})\citenamefont {Masjuan}, \citenamefont
  {Arriola},\ and\ \citenamefont {Broniowski}}]{Masjuan:2012gc}%
  \BibitemOpen
  \bibfield  {author} {\bibinfo {author} {\bibfnamefont {P.}~\bibnamefont
  {Masjuan}}, \bibinfo {author} {\bibfnamefont {E.~R.}\ \bibnamefont
  {Arriola}}, \ and\ \bibinfo {author} {\bibfnamefont {W.}~\bibnamefont
  {Broniowski}},\ }\href {\doibase 10.1103/PhysRevD.85.094006} {\bibfield
  {journal} {\bibinfo  {journal} {Phys.Rev.}\ }\textbf {\bibinfo {volume}
  {D85}},\ \bibinfo {pages} {094006} (\bibinfo {year} {2012}{\natexlab{a}})},\
  \Eprint {http://arxiv.org/abs/1203.4782} {arXiv:1203.4782 [hep-ph]}
  \BibitemShut {NoStop}%
%%CITATION = ARXIV:1203.4782;%%
\bibitem [{\citenamefont {Masjuan}\ \emph
  {et~al.}(2012{\natexlab{b}})\citenamefont {Masjuan}, \citenamefont
  {Arriola},\ and\ \citenamefont {Broniowski}}]{Masjuan:2012yw}%
  \BibitemOpen
  \bibfield  {author} {\bibinfo {author} {\bibfnamefont {P.}~\bibnamefont
  {Masjuan}}, \bibinfo {author} {\bibfnamefont {E.}~\bibnamefont {Arriola}}, \
  and\ \bibinfo {author} {\bibfnamefont {W.}~\bibnamefont {Broniowski}},\
  }\href@noop {} {\  (\bibinfo {year} {2012}{\natexlab{b}})},\ \Eprint
  {http://arxiv.org/abs/1208.4477} {arXiv:1208.4477 [hep-ph]} \BibitemShut
  {NoStop}%
%%CITATION = ARXIV:1208.4477;%%
\bibitem [{\citenamefont {'t~Hooft}(1974)}]{'tHooft:1973jz}%
  \BibitemOpen
  \bibfield  {author} {\bibinfo {author} {\bibfnamefont {G.}~\bibnamefont
  {'t~Hooft}},\ }\href {\doibase 10.1016/0550-3213(74)90154-0} {\bibfield
  {journal} {\bibinfo  {journal} {Nucl. Phys.}\ }\textbf {\bibinfo {volume}
  {B72}},\ \bibinfo {pages} {461} (\bibinfo {year} {1974})}\BibitemShut
  {NoStop}%
%%CITATION = NUPHA,B72,461;%%
\bibitem [{\citenamefont {Witten}(1979)}]{Witten:1979kh}%
  \BibitemOpen
  \bibfield  {author} {\bibinfo {author} {\bibfnamefont {E.}~\bibnamefont
  {Witten}},\ }\href {\doibase 10.1016/0550-3213(79)90232-3} {\bibfield
  {journal} {\bibinfo  {journal} {Nucl. Phys.}\ }\textbf {\bibinfo {volume}
  {B160}},\ \bibinfo {pages} {57} (\bibinfo {year} {1979})}\BibitemShut
  {NoStop}%
%%CITATION = NUPHA,B160,57;%%
\bibitem [{\citenamefont {Knecht}\ and\ \citenamefont
  {de~Rafael}(1998)}]{Knecht:1997ts}%
  \BibitemOpen
  \bibfield  {author} {\bibinfo {author} {\bibfnamefont {M.}~\bibnamefont
  {Knecht}}\ and\ \bibinfo {author} {\bibfnamefont {E.}~\bibnamefont
  {de~Rafael}},\ }\href@noop {} {\bibfield  {journal} {\bibinfo  {journal}
  {Phys. Lett.}\ }\textbf {\bibinfo {volume} {B424}},\ \bibinfo {pages} {335}
  (\bibinfo {year} {1998})},\ \Eprint {http://arxiv.org/abs/hep-ph/9712457}
  {hep-ph/9712457} \BibitemShut {NoStop}%
%%CITATION = HEP-PH 9712457;%%
\bibitem [{\citenamefont {Peris}\ \emph {et~al.}(1998)\citenamefont {Peris},
  \citenamefont {Perrottet},\ and\ \citenamefont {de~Rafael}}]{Peris:1998nj}%
  \BibitemOpen
  \bibfield  {author} {\bibinfo {author} {\bibfnamefont {S.}~\bibnamefont
  {Peris}}, \bibinfo {author} {\bibfnamefont {M.}~\bibnamefont {Perrottet}}, \
  and\ \bibinfo {author} {\bibfnamefont {E.}~\bibnamefont {de~Rafael}},\
  }\href@noop {} {\bibfield  {journal} {\bibinfo  {journal} {JHEP}\ }\textbf
  {\bibinfo {volume} {9805}},\ \bibinfo {pages} {011} (\bibinfo {year}
  {1998})},\ \Eprint {http://arxiv.org/abs/hep-ph/9805442}
  {arXiv:hep-ph/9805442 [hep-ph]} \BibitemShut {NoStop}%
%%CITATION = HEP-PH/9805442;%%
\bibitem [{\citenamefont {Pich}(2002)}]{Pich:2002xy}%
  \BibitemOpen
  \bibfield  {author} {\bibinfo {author} {\bibfnamefont {A.}~\bibnamefont
  {Pich}},\ }\href@noop {} {\  (\bibinfo {year} {2002})},\ \Eprint
  {http://arxiv.org/abs/hep-ph/0205030} {arXiv:hep-ph/0205030} \BibitemShut
  {NoStop}%
%%CITATION = HEP-PH/0205030;%%
\bibitem [{\citenamefont {Masjuan}\ and\ \citenamefont
  {Peris}(2007)}]{Masjuan:2007ay}%
  \BibitemOpen
  \bibfield  {author} {\bibinfo {author} {\bibfnamefont {P.}~\bibnamefont
  {Masjuan}}\ and\ \bibinfo {author} {\bibfnamefont {S.}~\bibnamefont
  {Peris}},\ }\href {\doibase 10.1088/1126-6708/2007/05/040} {\bibfield
  {journal} {\bibinfo  {journal} {JHEP}\ }\textbf {\bibinfo {volume} {0705}},\
  \bibinfo {pages} {040} (\bibinfo {year} {2007})},\ \Eprint
  {http://arxiv.org/abs/0704.1247} {arXiv:0704.1247 [hep-ph]} \BibitemShut
  {NoStop}%
%%CITATION = ARXIV:0704.1247;%%
\bibitem [{\citenamefont {Masjuan}\ and\ \citenamefont
  {Peris}(2008)}]{Masjuan:2008fr}%
  \BibitemOpen
  \bibfield  {author} {\bibinfo {author} {\bibfnamefont {P.}~\bibnamefont
  {Masjuan}}\ and\ \bibinfo {author} {\bibfnamefont {S.}~\bibnamefont
  {Peris}},\ }\href {\doibase 10.1016/j.physletb.2008.03.040} {\bibfield
  {journal} {\bibinfo  {journal} {Phys.Lett.}\ }\textbf {\bibinfo {volume}
  {B663}},\ \bibinfo {pages} {61} (\bibinfo {year} {2008})},\ \Eprint
  {http://arxiv.org/abs/0801.3558} {arXiv:0801.3558 [hep-ph]} \BibitemShut
  {NoStop}%
%%CITATION = ARXIV:0801.3558;%%
\bibitem [{\citenamefont {Broniowski}\ and\ \citenamefont
  {Arriola}(2009)}]{Broniowski:2009zh}%
  \BibitemOpen
  \bibfield  {author} {\bibinfo {author} {\bibfnamefont {W.}~\bibnamefont
  {Broniowski}}\ and\ \bibinfo {author} {\bibfnamefont {E.~R.}\ \bibnamefont
  {Arriola}},\ }\href {\doibase 10.1103/PhysRevD.79.057501} {\bibfield
  {journal} {\bibinfo  {journal} {Phys. Rev.}\ }\textbf {\bibinfo {volume}
  {D79}},\ \bibinfo {pages} {057501} (\bibinfo {year} {2009})},\ \Eprint
  {http://arxiv.org/abs/0901.3336} {arXiv:0901.3336 [hep-ph]} \BibitemShut
  {NoStop}%
%%CITATION = 0901.3336;%%
\bibitem [{\citenamefont {Rosell}\ \emph {et~al.}(2004)\citenamefont {Rosell},
  \citenamefont {Sanz-Cillero},\ and\ \citenamefont {Pich}}]{Rosell:2004mn}%
  \BibitemOpen
  \bibfield  {author} {\bibinfo {author} {\bibfnamefont {I.}~\bibnamefont
  {Rosell}}, \bibinfo {author} {\bibfnamefont {J.~J.}\ \bibnamefont
  {Sanz-Cillero}}, \ and\ \bibinfo {author} {\bibfnamefont {A.}~\bibnamefont
  {Pich}},\ }\href {\doibase 10.1088/1126-6708/2004/08/042} {\bibfield
  {journal} {\bibinfo  {journal} {JHEP}\ }\textbf {\bibinfo {volume} {0408}},\
  \bibinfo {pages} {042} (\bibinfo {year} {2004})},\ \Eprint
  {http://arxiv.org/abs/hep-ph/0407240} {arXiv:hep-ph/0407240 [hep-ph]}
  \BibitemShut {NoStop}%
\bibitem [{\citenamefont {Rosell}\ \emph {et~al.}(2005)\citenamefont {Rosell},
  \citenamefont {Ruiz-Femenia},\ and\ \citenamefont
  {Portoles}}]{Rosell:2005ai}%
  \BibitemOpen
  \bibfield  {author} {\bibinfo {author} {\bibfnamefont {I.}~\bibnamefont
  {Rosell}}, \bibinfo {author} {\bibfnamefont {P.}~\bibnamefont
  {Ruiz-Femenia}}, \ and\ \bibinfo {author} {\bibfnamefont {J.}~\bibnamefont
  {Portoles}},\ }\href {\doibase 10.1088/1126-6708/2005/12/020} {\bibfield
  {journal} {\bibinfo  {journal} {JHEP}\ }\textbf {\bibinfo {volume} {0512}},\
  \bibinfo {pages} {020} (\bibinfo {year} {2005})},\ \Eprint
  {http://arxiv.org/abs/hep-ph/0510041} {arXiv:hep-ph/0510041 [hep-ph]}
  \BibitemShut {NoStop}%
\bibitem [{\citenamefont {Rosell}\ \emph {et~al.}(2007)\citenamefont {Rosell},
  \citenamefont {Sanz-Cillero},\ and\ \citenamefont {Pich}}]{Rosell:2006dt}%
  \BibitemOpen
  \bibfield  {author} {\bibinfo {author} {\bibfnamefont {I.}~\bibnamefont
  {Rosell}}, \bibinfo {author} {\bibfnamefont {J.~J.}\ \bibnamefont
  {Sanz-Cillero}}, \ and\ \bibinfo {author} {\bibfnamefont {A.}~\bibnamefont
  {Pich}},\ }\href {\doibase 10.1088/1126-6708/2007/01/039} {\bibfield
  {journal} {\bibinfo  {journal} {JHEP}\ }\textbf {\bibinfo {volume} {0701}},\
  \bibinfo {pages} {039} (\bibinfo {year} {2007})},\ \Eprint
  {http://arxiv.org/abs/hep-ph/0610290} {arXiv:hep-ph/0610290 [hep-ph]}
  \BibitemShut {NoStop}%
\bibitem [{\citenamefont {Pich}\ \emph {et~al.}(2011)\citenamefont {Pich},
  \citenamefont {Rosell},\ and\ \citenamefont {Sanz-Cillero}}]{Pich:2010sm}%
  \BibitemOpen
  \bibfield  {author} {\bibinfo {author} {\bibfnamefont {A.}~\bibnamefont
  {Pich}}, \bibinfo {author} {\bibfnamefont {I.}~\bibnamefont {Rosell}}, \ and\
  \bibinfo {author} {\bibfnamefont {J.~J.}\ \bibnamefont {Sanz-Cillero}},\
  }\href {\doibase 10.1007/JHEP02(2011)109} {\bibfield  {journal} {\bibinfo
  {journal} {JHEP}\ }\textbf {\bibinfo {volume} {1102}},\ \bibinfo {pages}
  {109} (\bibinfo {year} {2011})},\ \Eprint {http://arxiv.org/abs/1011.5771
  [hep-ph]} {arXiv:1011.5771 [hep-ph] [hep-ph]} \BibitemShut {NoStop}%
\bibitem [{\citenamefont {Pich}\ \emph {et~al.}(2008)\citenamefont {Pich},
  \citenamefont {Rosell},\ and\ \citenamefont {Sanz-Cillero}}]{Pich:2008jm}%
  \BibitemOpen
  \bibfield  {author} {\bibinfo {author} {\bibfnamefont {A.}~\bibnamefont
  {Pich}}, \bibinfo {author} {\bibfnamefont {I.}~\bibnamefont {Rosell}}, \ and\
  \bibinfo {author} {\bibfnamefont {J.~J.}\ \bibnamefont {Sanz-Cillero}},\
  }\href {\doibase 10.1088/1126-6708/2008/07/014} {\bibfield  {journal}
  {\bibinfo  {journal} {JHEP}\ }\textbf {\bibinfo {volume} {0807}},\ \bibinfo
  {pages} {014} (\bibinfo {year} {2008})},\ \Eprint
  {http://arxiv.org/abs/0803.1567 [hep-ph]} {arXiv:0803.1567 [hep-ph] [hep-ph]}
  \BibitemShut {NoStop}%
\bibitem [{\citenamefont {Golterman}\ and\ \citenamefont
  {Peris}(2001)}]{Golterman:2001nk}%
  \BibitemOpen
  \bibfield  {author} {\bibinfo {author} {\bibfnamefont {M.}~\bibnamefont
  {Golterman}}\ and\ \bibinfo {author} {\bibfnamefont {S.}~\bibnamefont
  {Peris}},\ }\href@noop {} {\bibfield  {journal} {\bibinfo  {journal} {JHEP}\
  }\textbf {\bibinfo {volume} {01}},\ \bibinfo {pages} {028} (\bibinfo {year}
  {2001})},\ \Eprint {http://arxiv.org/abs/hep-ph/0101098}
  {arXiv:hep-ph/0101098} \BibitemShut {NoStop}%
%%CITATION = HEP-PH/0101098;%%
\bibitem [{\citenamefont {Beane}(2001)}]{Beane:2001uj}%
  \BibitemOpen
  \bibfield  {author} {\bibinfo {author} {\bibfnamefont {S.~R.}\ \bibnamefont
  {Beane}},\ }\href {\doibase 10.1103/PhysRevD.64.116010} {\bibfield  {journal}
  {\bibinfo  {journal} {Phys. Rev.}\ }\textbf {\bibinfo {volume} {D64}},\
  \bibinfo {pages} {116010} (\bibinfo {year} {2001})},\ \Eprint
  {http://arxiv.org/abs/hep-ph/0106022} {arXiv:hep-ph/0106022} \BibitemShut
  {NoStop}%
%%CITATION = HEP-PH/0106022;%%
\bibitem [{\citenamefont {Afonin}\ \emph {et~al.}(2004)\citenamefont {Afonin},
  \citenamefont {Andrianov}, \citenamefont {Andrianov},\ and\ \citenamefont
  {Espriu}}]{Afonin:2004yb}%
  \BibitemOpen
  \bibfield  {author} {\bibinfo {author} {\bibfnamefont {S.~S.}\ \bibnamefont
  {Afonin}}, \bibinfo {author} {\bibfnamefont {A.~A.}\ \bibnamefont
  {Andrianov}}, \bibinfo {author} {\bibfnamefont {V.~A.}\ \bibnamefont
  {Andrianov}}, \ and\ \bibinfo {author} {\bibfnamefont {D.}~\bibnamefont
  {Espriu}},\ }\href {\doibase 10.1088/1126-6708/2004/04/039} {\bibfield
  {journal} {\bibinfo  {journal} {JHEP}\ }\textbf {\bibinfo {volume} {04}},\
  \bibinfo {pages} {039} (\bibinfo {year} {2004})},\ \Eprint
  {http://arxiv.org/abs/hep-ph/0403268} {arXiv:hep-ph/0403268} \BibitemShut
  {NoStop}%
%%CITATION = HEP-PH/0403268;%%
\bibitem [{\citenamefont {Ruiz~Arriola}\ and\ \citenamefont
  {Broniowski}(2006{\natexlab{a}})}]{RuizArriola:2006gq}%
  \BibitemOpen
  \bibfield  {author} {\bibinfo {author} {\bibfnamefont {E.}~\bibnamefont
  {Ruiz~Arriola}}\ and\ \bibinfo {author} {\bibfnamefont {W.}~\bibnamefont
  {Broniowski}},\ }\href {\doibase 10.1103/PhysRevD.73.097502} {\bibfield
  {journal} {\bibinfo  {journal} {Phys. Rev.}\ }\textbf {\bibinfo {volume}
  {D73}},\ \bibinfo {pages} {097502} (\bibinfo {year} {2006}{\natexlab{a}})},\
  \Eprint {http://arxiv.org/abs/hep-ph/0603263} {arXiv:hep-ph/0603263}
  \BibitemShut {NoStop}%
%%CITATION = HEP-PH/0603263;%%
\bibitem [{\citenamefont {Arriola}\ and\ \citenamefont
  {Broniowski}(2007)}]{Arriola:2006sv}%
  \BibitemOpen
  \bibfield  {author} {\bibinfo {author} {\bibfnamefont {E.~R.}\ \bibnamefont
  {Arriola}}\ and\ \bibinfo {author} {\bibfnamefont {W.}~\bibnamefont
  {Broniowski}},\ }\href {\doibase 10.1140/epja/i2006-10184-7} {\bibfield
  {journal} {\bibinfo  {journal} {Eur. Phys. J.}\ }\textbf {\bibinfo {volume}
  {A31}},\ \bibinfo {pages} {739} (\bibinfo {year} {2007})},\ \Eprint
  {http://arxiv.org/abs/hep-ph/0609266} {arXiv:hep-ph/0609266} \BibitemShut
  {NoStop}%
%%CITATION = HEP-PH/0609266;%%
\bibitem [{\citenamefont {Masjuan~Queralt}(2010)}]{Queralt:2010sv}%
  \BibitemOpen
  \bibfield  {author} {\bibinfo {author} {\bibfnamefont {P.}~\bibnamefont
  {Masjuan~Queralt}},\ }\href@noop {} {\  (\bibinfo {year} {2010})},\ \Eprint
  {http://arxiv.org/abs/1005.5683} {arXiv:1005.5683 [hep-ph]} \BibitemShut
  {NoStop}%
%%CITATION = ARXIV:1005.5683;%%
\bibitem [{\citenamefont {Anisovich}\ \emph {et~al.}(2000)\citenamefont
  {Anisovich}, \citenamefont {Anisovich},\ and\ \citenamefont
  {Sarantsev}}]{Anisovich:2000kxa}%
  \BibitemOpen
  \bibfield  {author} {\bibinfo {author} {\bibfnamefont {A.~V.}\ \bibnamefont
  {Anisovich}}, \bibinfo {author} {\bibfnamefont {V.~V.}\ \bibnamefont
  {Anisovich}}, \ and\ \bibinfo {author} {\bibfnamefont {A.~V.}\ \bibnamefont
  {Sarantsev}},\ }\href {\doibase 10.1103/PhysRevD.62.051502} {\bibfield
  {journal} {\bibinfo  {journal} {Phys. Rev.}\ }\textbf {\bibinfo {volume}
  {D62}},\ \bibinfo {pages} {051502} (\bibinfo {year} {2000})},\ \Eprint
  {http://arxiv.org/abs/hep-ph/0003113} {arXiv:hep-ph/0003113} \BibitemShut
  {NoStop}%
%%CITATION = HEP-PH/0003113;%%
\bibitem [{\citenamefont {Dominguez}(2001)}]{Dominguez:2001zu}%
  \BibitemOpen
  \bibfield  {author} {\bibinfo {author} {\bibfnamefont {C.~A.}\ \bibnamefont
  {Dominguez}},\ }\href {\doibase 10.1016/S0370-2693(01)00576-7} {\bibfield
  {journal} {\bibinfo  {journal} {Phys. Lett.}\ }\textbf {\bibinfo {volume}
  {B512}},\ \bibinfo {pages} {331} (\bibinfo {year} {2001})},\ \Eprint
  {http://arxiv.org/abs/hep-ph/0102190} {arXiv:hep-ph/0102190} \BibitemShut
  {NoStop}%
%%CITATION = HEP-PH/0102190;%%
\bibitem [{\citenamefont {Ruiz~Arriola}\ and\ \citenamefont
  {Broniowski}(2006{\natexlab{b}})}]{RuizArriola:2006ii}%
  \BibitemOpen
  \bibfield  {author} {\bibinfo {author} {\bibfnamefont {E.}~\bibnamefont
  {Ruiz~Arriola}}\ and\ \bibinfo {author} {\bibfnamefont {W.}~\bibnamefont
  {Broniowski}},\ }\href {\doibase 10.1103/PhysRevD.74.034008} {\bibfield
  {journal} {\bibinfo  {journal} {Phys. Rev.}\ }\textbf {\bibinfo {volume}
  {D74}},\ \bibinfo {pages} {034008} (\bibinfo {year} {2006}{\natexlab{b}})},\
  \Eprint {http://arxiv.org/abs/hep-ph/0605318} {arXiv:hep-ph/0605318}
  \BibitemShut {NoStop}%
%%CITATION = HEP-PH/0605318;%%
\bibitem [{\citenamefont {Ruiz~Arriola}\ and\ \citenamefont
  {Broniowski}(2008)}]{RuizArriola:2008sq}%
  \BibitemOpen
  \bibfield  {author} {\bibinfo {author} {\bibfnamefont {E.}~\bibnamefont
  {Ruiz~Arriola}}\ and\ \bibinfo {author} {\bibfnamefont {W.}~\bibnamefont
  {Broniowski}},\ }\href {\doibase 10.1103/PhysRevD.78.034031} {\bibfield
  {journal} {\bibinfo  {journal} {Phys. Rev.}\ }\textbf {\bibinfo {volume}
  {D78}},\ \bibinfo {pages} {034031} (\bibinfo {year} {2008})},\ \Eprint
  {http://arxiv.org/abs/0807.3488} {arXiv:0807.3488 [hep-ph]} \BibitemShut
  {NoStop}%
%%CITATION = 0807.3488;%%
\bibitem [{\citenamefont {Bijnens}\ \emph {et~al.}(2003)\citenamefont
  {Bijnens}, \citenamefont {Gamiz}, \citenamefont {Lipartia},\ and\
  \citenamefont {Prades}}]{Bijnens:2003rc}%
  \BibitemOpen
  \bibfield  {author} {\bibinfo {author} {\bibfnamefont {J.}~\bibnamefont
  {Bijnens}}, \bibinfo {author} {\bibfnamefont {E.}~\bibnamefont {Gamiz}},
  \bibinfo {author} {\bibfnamefont {E.}~\bibnamefont {Lipartia}}, \ and\
  \bibinfo {author} {\bibfnamefont {J.}~\bibnamefont {Prades}},\ }\href@noop {}
  {\bibfield  {journal} {\bibinfo  {journal} {JHEP}\ }\textbf {\bibinfo
  {volume} {04}},\ \bibinfo {pages} {055} (\bibinfo {year} {2003})},\ \Eprint
  {http://arxiv.org/abs/hep-ph/0304222} {arXiv:hep-ph/0304222} \BibitemShut
  {NoStop}%
%%CITATION = HEP-PH/0304222;%%
\bibitem [{\citenamefont {Ananthanarayan}\ \emph {et~al.}(2012)\citenamefont
  {Ananthanarayan}, \citenamefont {Caprini},\ and\ \citenamefont
  {Imsong}}]{Ananthanarayan:2012tn}%
  \BibitemOpen
  \bibfield  {author} {\bibinfo {author} {\bibfnamefont {B.}~\bibnamefont
  {Ananthanarayan}}, \bibinfo {author} {\bibfnamefont {I.}~\bibnamefont
  {Caprini}}, \ and\ \bibinfo {author} {\bibfnamefont {I.~S.}\ \bibnamefont
  {Imsong}},\ }\href@noop {} {\bibfield  {journal} {\bibinfo  {journal}
  {Phys.Rev.}\ }\textbf {\bibinfo {volume} {D85}},\ \bibinfo {pages} {096006}
  (\bibinfo {year} {2012})},\ \Eprint {http://arxiv.org/abs/1203.5398}
  {arXiv:1203.5398 [hep-ph]} \BibitemShut {NoStop}%
%%CITATION = ARXIV:1203.5398;%%
\bibitem [{\citenamefont {Hong}\ \emph {et~al.}(2008)\citenamefont {Hong},
  \citenamefont {Rho}, \citenamefont {Yee},\ and\ \citenamefont
  {Yi}}]{Hong:2007dq}%
  \BibitemOpen
  \bibfield  {author} {\bibinfo {author} {\bibfnamefont {D.~K.}\ \bibnamefont
  {Hong}}, \bibinfo {author} {\bibfnamefont {M.}~\bibnamefont {Rho}}, \bibinfo
  {author} {\bibfnamefont {H.-U.}\ \bibnamefont {Yee}}, \ and\ \bibinfo
  {author} {\bibfnamefont {P.}~\bibnamefont {Yi}},\ }\href {\doibase
  10.1103/PhysRevD.77.014030} {\bibfield  {journal} {\bibinfo  {journal} {Phys.
  Rev.}\ }\textbf {\bibinfo {volume} {D77}},\ \bibinfo {pages} {014030}
  (\bibinfo {year} {2008})},\ \Eprint {http://arxiv.org/abs/0710.4615}
  {arXiv:0710.4615 [hep-ph]} \BibitemShut {NoStop}%
%%CITATION = 0710.4615;%%
\bibitem [{\citenamefont {Harada}\ \emph {et~al.}(2010)\citenamefont {Harada},
  \citenamefont {Matsuzaki},\ and\ \citenamefont {Yamawaki}}]{Harada:2010cn}%
  \BibitemOpen
  \bibfield  {author} {\bibinfo {author} {\bibfnamefont {M.}~\bibnamefont
  {Harada}}, \bibinfo {author} {\bibfnamefont {S.}~\bibnamefont {Matsuzaki}}, \
  and\ \bibinfo {author} {\bibfnamefont {K.}~\bibnamefont {Yamawaki}},\ }\href
  {\doibase 10.1103/PhysRevD.82.076010} {\bibfield  {journal} {\bibinfo
  {journal} {Phys. Rev.}\ }\textbf {\bibinfo {volume} {D82}},\ \bibinfo {pages}
  {076010} (\bibinfo {year} {2010})},\ \Eprint {http://arxiv.org/abs/1007.4715}
  {arXiv:1007.4715 [hep-ph]} \BibitemShut {NoStop}%
%%CITATION = 1007.4715;%%
\bibitem [{\citenamefont {Belushkin}\ \emph {et~al.}(2007)\citenamefont
  {Belushkin}, \citenamefont {Hammer},\ and\ \citenamefont
  {Meissner}}]{Belushkin:2006qa}%
  \BibitemOpen
  \bibfield  {author} {\bibinfo {author} {\bibfnamefont {M.~A.}\ \bibnamefont
  {Belushkin}}, \bibinfo {author} {\bibfnamefont {H.~W.}\ \bibnamefont
  {Hammer}}, \ and\ \bibinfo {author} {\bibfnamefont {U.~G.}\ \bibnamefont
  {Meissner}},\ }\href {\doibase 10.1103/PhysRevC.75.035202} {\bibfield
  {journal} {\bibinfo  {journal} {Phys. Rev.}\ }\textbf {\bibinfo {volume}
  {C75}},\ \bibinfo {pages} {035202} (\bibinfo {year} {2007})},\ \Eprint
  {http://arxiv.org/abs/hep-ph/0608337} {arXiv:hep-ph/0608337} \BibitemShut
  {NoStop}%
%%CITATION = HEP-PH/0608337;%%
\bibitem [{\citenamefont {Portoles}(2010)}]{Portoles:2010yt}%
  \BibitemOpen
  \bibfield  {author} {\bibinfo {author} {\bibfnamefont {J.}~\bibnamefont
  {Portoles}},\ }\href {\doibase 10.1063/1.3541978} {\bibfield  {journal}
  {\bibinfo  {journal} {AIP Conf.Proc.}\ }\textbf {\bibinfo {volume} {1322}},\
  \bibinfo {pages} {178} (\bibinfo {year} {2010})},\ \Eprint
  {http://arxiv.org/abs/1010.3360} {arXiv:1010.3360 [hep-ph]} \BibitemShut
  {NoStop}%
%%CITATION = ARXIV:1010.3360;%%
\bibitem [{\citenamefont {Teper}(2009)}]{Teper:2009uf}%
  \BibitemOpen
  \bibfield  {author} {\bibinfo {author} {\bibfnamefont {M.}~\bibnamefont
  {Teper}},\ }\href@noop {} {\bibfield  {journal} {\bibinfo  {journal} {Acta
  Phys. Polon.}\ }\textbf {\bibinfo {volume} {B40}},\ \bibinfo {pages} {3249}
  (\bibinfo {year} {2009})},\ \Eprint {http://arxiv.org/abs/0912.3339}
  {arXiv:0912.3339 [hep-lat]} \BibitemShut {NoStop}%
%%CITATION = 0912.3339;%%
\bibitem [{\citenamefont {Kampf}\ \emph {et~al.}(2010)\citenamefont {Kampf},
  \citenamefont {Novotny},\ and\ \citenamefont {Trnka}}]{Kampf:2009jh}%
  \BibitemOpen
  \bibfield  {author} {\bibinfo {author} {\bibfnamefont {K.}~\bibnamefont
  {Kampf}}, \bibinfo {author} {\bibfnamefont {J.}~\bibnamefont {Novotny}}, \
  and\ \bibinfo {author} {\bibfnamefont {J.}~\bibnamefont {Trnka}},\ }\href
  {\doibase 10.1103/PhysRevD.81.116004} {\bibfield  {journal} {\bibinfo
  {journal} {Phys.Rev.}\ }\textbf {\bibinfo {volume} {D81}},\ \bibinfo {pages}
  {116004} (\bibinfo {year} {2010})},\ \Eprint {http://arxiv.org/abs/0912.5289}
  {arXiv:0912.5289 [hep-ph]} \BibitemShut {NoStop}%
%%CITATION = ARXIV:0912.5289;%%
\bibitem [{\citenamefont {Garcia-Martin}\ \emph {et~al.}(2011)\citenamefont
  {Garcia-Martin}, \citenamefont {Kaminski}, \citenamefont {Pelaez},
  \citenamefont {Ruiz~de Elvira},\ and\ \citenamefont
  {Yndurain}}]{GarciaMartin:2011cn}%
  \BibitemOpen
  \bibfield  {author} {\bibinfo {author} {\bibfnamefont {R.}~\bibnamefont
  {Garcia-Martin}}, \bibinfo {author} {\bibfnamefont {R.}~\bibnamefont
  {Kaminski}}, \bibinfo {author} {\bibfnamefont {J.}~\bibnamefont {Pelaez}},
  \bibinfo {author} {\bibfnamefont {J.}~\bibnamefont {Ruiz~de Elvira}}, \ and\
  \bibinfo {author} {\bibfnamefont {F.}~\bibnamefont {Yndurain}},\ }\href
  {\doibase 10.1103/PhysRevD.83.074004} {\bibfield  {journal} {\bibinfo
  {journal} {Phys.Rev.}\ }\textbf {\bibinfo {volume} {D83}},\ \bibinfo {pages}
  {074004} (\bibinfo {year} {2011})},\ \Eprint {http://arxiv.org/abs/1102.2183}
  {arXiv:1102.2183 [hep-ph]} \BibitemShut {NoStop}%
%%CITATION = ARXIV:1102.2183;%%
\bibitem [{\citenamefont {Gounaris}\ and\ \citenamefont
  {Sakurai}(1968)}]{Gounaris:1968mw}%
  \BibitemOpen
  \bibfield  {author} {\bibinfo {author} {\bibfnamefont {G.~J.}\ \bibnamefont
  {Gounaris}}\ and\ \bibinfo {author} {\bibfnamefont {J.~J.}\ \bibnamefont
  {Sakurai}},\ }\href {\doibase 10.1103/PhysRevLett.21.244} {\bibfield
  {journal} {\bibinfo  {journal} {Phys. Rev. Lett.}\ }\textbf {\bibinfo
  {volume} {21}},\ \bibinfo {pages} {244} (\bibinfo {year} {1968})}\BibitemShut
  {NoStop}%
%%CITATION = PRLTA,21,244;%%
\bibitem [{\citenamefont {Nieves}\ and\ \citenamefont
  {Ruiz~Arriola}(2000)}]{Nieves:1999bx}%
  \BibitemOpen
  \bibfield  {author} {\bibinfo {author} {\bibfnamefont {J.}~\bibnamefont
  {Nieves}}\ and\ \bibinfo {author} {\bibfnamefont {E.}~\bibnamefont
  {Ruiz~Arriola}},\ }\href {\doibase 10.1016/S0375-9474(00)00321-3} {\bibfield
  {journal} {\bibinfo  {journal} {Nucl. Phys.}\ }\textbf {\bibinfo {volume}
  {A679}},\ \bibinfo {pages} {57} (\bibinfo {year} {2000})},\ \Eprint
  {http://arxiv.org/abs/hep-ph/9907469} {arXiv:hep-ph/9907469} \BibitemShut
  {NoStop}%
%%CITATION = HEP-PH/9907469;%%
\bibitem [{\citenamefont {Brown}\ \emph {et~al.}(1973)\citenamefont {Brown},
  \citenamefont {Canizares}, \citenamefont {Cooper}, \citenamefont {Eisner},
  \citenamefont {Feldmann} \emph {et~al.}}]{Brown:1973wr}%
  \BibitemOpen
  \bibfield  {author} {\bibinfo {author} {\bibfnamefont {C.}~\bibnamefont
  {Brown}}, \bibinfo {author} {\bibfnamefont {C.}~\bibnamefont {Canizares}},
  \bibinfo {author} {\bibfnamefont {W.}~\bibnamefont {Cooper}}, \bibinfo
  {author} {\bibfnamefont {A.}~\bibnamefont {Eisner}}, \bibinfo {author}
  {\bibfnamefont {G.}~\bibnamefont {Feldmann}},  \emph {et~al.},\ }\href
  {\doibase 10.1103/PhysRevD.8.92} {\bibfield  {journal} {\bibinfo  {journal}
  {Phys.Rev.}\ }\textbf {\bibinfo {volume} {D8}},\ \bibinfo {pages} {92}
  (\bibinfo {year} {1973})}\BibitemShut {NoStop}%
%%CITATION = PHRVA,D8,92;%%
\bibitem [{\citenamefont {Bebek}\ \emph {et~al.}(1974)\citenamefont {Bebek},
  \citenamefont {Brown}, \citenamefont {Herzlinger}, \citenamefont {Holmes},
  \citenamefont {Lichtenstein} \emph {et~al.}}]{Bebek:1974iz}%
  \BibitemOpen
  \bibfield  {author} {\bibinfo {author} {\bibfnamefont {C.}~\bibnamefont
  {Bebek}}, \bibinfo {author} {\bibfnamefont {C.}~\bibnamefont {Brown}},
  \bibinfo {author} {\bibfnamefont {M.}~\bibnamefont {Herzlinger}}, \bibinfo
  {author} {\bibfnamefont {S.~D.}\ \bibnamefont {Holmes}}, \bibinfo {author}
  {\bibfnamefont {C.}~\bibnamefont {Lichtenstein}},  \emph {et~al.},\ }\href
  {\doibase 10.1103/PhysRevD.9.1229} {\bibfield  {journal} {\bibinfo  {journal}
  {Phys.Rev.}\ }\textbf {\bibinfo {volume} {D9}},\ \bibinfo {pages} {1229}
  (\bibinfo {year} {1974})}\BibitemShut {NoStop}%
%%CITATION = PHRVA,D9,1229;%%
\bibitem [{\citenamefont {Bebek}\ \emph {et~al.}(1976)\citenamefont {Bebek}
  \emph {et~al.}}]{Bebek:1974ww}%
  \BibitemOpen
  \bibfield  {author} {\bibinfo {author} {\bibfnamefont {C.~J.}\ \bibnamefont
  {Bebek}} \emph {et~al.},\ }\href {\doibase 10.1103/PhysRevD.13.25} {\bibfield
   {journal} {\bibinfo  {journal} {Phys. Rev.}\ }\textbf {\bibinfo {volume}
  {D13}},\ \bibinfo {pages} {25} (\bibinfo {year} {1976})}\BibitemShut
  {NoStop}%
%%CITATION = PHRVA,D13,25;%%
\bibitem [{\citenamefont {Bebek}\ \emph {et~al.}(1978)\citenamefont {Bebek}
  \emph {et~al.}}]{Bebek:1977pe}%
  \BibitemOpen
  \bibfield  {author} {\bibinfo {author} {\bibfnamefont {C.~J.}\ \bibnamefont
  {Bebek}} \emph {et~al.},\ }\href {\doibase 10.1103/PhysRevD.17.1693}
  {\bibfield  {journal} {\bibinfo  {journal} {Phys. Rev.}\ }\textbf {\bibinfo
  {volume} {D17}},\ \bibinfo {pages} {1693} (\bibinfo {year}
  {1978})}\BibitemShut {NoStop}%
%%CITATION = PHRVA,D17,1693;%%
\bibitem [{\citenamefont {Dally}\ \emph {et~al.}(1977)\citenamefont {Dally},
  \citenamefont {Drickey}, \citenamefont {Hauptman}, \citenamefont {May},
  \citenamefont {Stork} \emph {et~al.}}]{Dally:1977vt}%
  \BibitemOpen
  \bibfield  {author} {\bibinfo {author} {\bibfnamefont {E.}~\bibnamefont
  {Dally}}, \bibinfo {author} {\bibfnamefont {D.~J.}\ \bibnamefont {Drickey}},
  \bibinfo {author} {\bibfnamefont {J.}~\bibnamefont {Hauptman}}, \bibinfo
  {author} {\bibfnamefont {C.}~\bibnamefont {May}}, \bibinfo {author}
  {\bibfnamefont {D.}~\bibnamefont {Stork}},  \emph {et~al.},\ }\href {\doibase
  10.1103/PhysRevLett.39.1176} {\bibfield  {journal} {\bibinfo  {journal}
  {Phys.Rev.Lett.}\ }\textbf {\bibinfo {volume} {39}},\ \bibinfo {pages} {1176}
  (\bibinfo {year} {1977})}\BibitemShut {NoStop}%
%%CITATION = PRLTA,39,1176;%%
\bibitem [{\citenamefont {Brauel}\ \emph {et~al.}(1979)\citenamefont {Brauel},
  \citenamefont {Canzler}, \citenamefont {Cords}, \citenamefont {Felst},
  \citenamefont {Grindhammer} \emph {et~al.}}]{Brauel:1979zk}%
  \BibitemOpen
  \bibfield  {author} {\bibinfo {author} {\bibfnamefont {P.}~\bibnamefont
  {Brauel}}, \bibinfo {author} {\bibfnamefont {T.}~\bibnamefont {Canzler}},
  \bibinfo {author} {\bibfnamefont {D.}~\bibnamefont {Cords}}, \bibinfo
  {author} {\bibfnamefont {R.}~\bibnamefont {Felst}}, \bibinfo {author}
  {\bibfnamefont {G.}~\bibnamefont {Grindhammer}},  \emph {et~al.},\
  }\href@noop {} {\bibfield  {journal} {\bibinfo  {journal} {Z.Phys.}\ }\textbf
  {\bibinfo {volume} {C3}},\ \bibinfo {pages} {101} (\bibinfo {year}
  {1979})}\BibitemShut {NoStop}%
\bibitem [{\citenamefont {Amendolia}\ \emph {et~al.}(1986)\citenamefont
  {Amendolia} \emph {et~al.}}]{Amendolia:1986wj}%
  \BibitemOpen
  \bibfield  {author} {\bibinfo {author} {\bibfnamefont {S.~R.}\ \bibnamefont
  {Amendolia}} \emph {et~al.} (\bibinfo {collaboration} {NA7}),\ }\href@noop {}
  {\bibfield  {journal} {\bibinfo  {journal} {Nucl. Phys.}\ }\textbf {\bibinfo
  {volume} {B277}},\ \bibinfo {pages} {168} (\bibinfo {year}
  {1986})}\BibitemShut {NoStop}%
%%CITATION = NUPHA,B277,168;%%
\bibitem [{\citenamefont {Volmer}\ \emph {et~al.}(2001)\citenamefont {Volmer}
  \emph {et~al.}}]{Volmer:2000ek}%
  \BibitemOpen
  \bibfield  {author} {\bibinfo {author} {\bibfnamefont {J.}~\bibnamefont
  {Volmer}} \emph {et~al.} (\bibinfo {collaboration} {Jefferson Lab F(pi)
  Collaboration}),\ }\href {\doibase 10.1103/PhysRevLett.86.1713} {\bibfield
  {journal} {\bibinfo  {journal} {Phys.Rev.Lett.}\ }\textbf {\bibinfo {volume}
  {86}},\ \bibinfo {pages} {1713} (\bibinfo {year} {2001})},\ \Eprint
  {http://arxiv.org/abs/nucl-ex/0010009} {arXiv:nucl-ex/0010009 [nucl-ex]}
  \BibitemShut {NoStop}%
%%CITATION = NUCL-EX/0010009;%%
\bibitem [{\citenamefont {Tadevosyan}\ \emph {et~al.}(2007)\citenamefont
  {Tadevosyan} \emph {et~al.}}]{Tadevosyan:2007yd}%
  \BibitemOpen
  \bibfield  {author} {\bibinfo {author} {\bibfnamefont {V.}~\bibnamefont
  {Tadevosyan}} \emph {et~al.} (\bibinfo {collaboration} {Jefferson Lab
  F(pi)}),\ }\href {\doibase 10.1103/PhysRevC.75.055205} {\bibfield  {journal}
  {\bibinfo  {journal} {Phys. Rev.}\ }\textbf {\bibinfo {volume} {C75}},\
  \bibinfo {pages} {055205} (\bibinfo {year} {2007})},\ \Eprint
  {http://arxiv.org/abs/nucl-ex/0607007} {arXiv:nucl-ex/0607007} \BibitemShut
  {NoStop}%
%%CITATION = NUCL-EX/0607007;%%
\bibitem [{\citenamefont {Horn}\ \emph {et~al.}(2006)\citenamefont {Horn} \emph
  {et~al.}}]{Horn:2006tm}%
  \BibitemOpen
  \bibfield  {author} {\bibinfo {author} {\bibfnamefont {T.}~\bibnamefont
  {Horn}} \emph {et~al.} (\bibinfo {collaboration} {Jefferson Lab F(pi)-2}),\
  }\href {\doibase 10.1103/PhysRevLett.97.192001} {\bibfield  {journal}
  {\bibinfo  {journal} {Phys. Rev. Lett.}\ }\textbf {\bibinfo {volume} {97}},\
  \bibinfo {pages} {192001} (\bibinfo {year} {2006})},\ \Eprint
  {http://arxiv.org/abs/nucl-ex/0607005} {arXiv:nucl-ex/0607005} \BibitemShut
  {NoStop}%
%%CITATION = NUCL-EX/0607005;%%
\bibitem [{\citenamefont {Horn}\ \emph {et~al.}(2008)\citenamefont {Horn},
  \citenamefont {Qian}, \citenamefont {Arrington}, \citenamefont {Asaturyan},
  \citenamefont {Benmokthar} \emph {et~al.}}]{Horn:2007ug}%
  \BibitemOpen
  \bibfield  {author} {\bibinfo {author} {\bibfnamefont {T.}~\bibnamefont
  {Horn}}, \bibinfo {author} {\bibfnamefont {X.}~\bibnamefont {Qian}}, \bibinfo
  {author} {\bibfnamefont {J.}~\bibnamefont {Arrington}}, \bibinfo {author}
  {\bibfnamefont {R.}~\bibnamefont {Asaturyan}}, \bibinfo {author}
  {\bibfnamefont {F.}~\bibnamefont {Benmokthar}},  \emph {et~al.},\ }\href
  {\doibase 10.1103/PhysRevC.78.058201} {\bibfield  {journal} {\bibinfo
  {journal} {Phys.Rev.}\ }\textbf {\bibinfo {volume} {C78}},\ \bibinfo {pages}
  {058201} (\bibinfo {year} {2008})},\ \Eprint {http://arxiv.org/abs/0707.1794}
  {arXiv:0707.1794 [nucl-ex]} \BibitemShut {NoStop}%
%%CITATION = ARXIV:0707.1794;%%
\bibitem [{\citenamefont {Behrend}\ \emph {et~al.}(1991)\citenamefont {Behrend}
  \emph {et~al.}}]{Behrend:1990sr}%
  \BibitemOpen
  \bibfield  {author} {\bibinfo {author} {\bibfnamefont {H.~J.}\ \bibnamefont
  {Behrend}} \emph {et~al.} (\bibinfo {collaboration} {CELLO}),\ }\href@noop {}
  {\bibfield  {journal} {\bibinfo  {journal} {Z. Phys.}\ }\textbf {\bibinfo
  {volume} {C49}},\ \bibinfo {pages} {401} (\bibinfo {year}
  {1991})}\BibitemShut {NoStop}%
%%CITATION = ZEPYA,C49,401;%%
\bibitem [{\citenamefont {Gronberg}\ \emph {et~al.}(1998)\citenamefont
  {Gronberg} \emph {et~al.}}]{Gronberg:1997fj}%
  \BibitemOpen
  \bibfield  {author} {\bibinfo {author} {\bibfnamefont {J.}~\bibnamefont
  {Gronberg}} \emph {et~al.} (\bibinfo {collaboration} {CLEO}),\ }\href@noop {}
  {\bibfield  {journal} {\bibinfo  {journal} {Phys. Rev.}\ }\textbf {\bibinfo
  {volume} {D57}},\ \bibinfo {pages} {33} (\bibinfo {year} {1998})},\ \Eprint
  {http://arxiv.org/abs/hep-ex/9707031} {hep-ex/9707031} \BibitemShut {NoStop}%
%%CITATION = HEP-EX 9707031;%%
\bibitem [{\citenamefont {Aubert}\ \emph {et~al.}(2009)\citenamefont {Aubert}
  \emph {et~al.}}]{Aubert:2009mc}%
  \BibitemOpen
  \bibfield  {author} {\bibinfo {author} {\bibfnamefont {B.}~\bibnamefont
  {Aubert}} \emph {et~al.} (\bibinfo {collaboration} {BABAR Collaboration}),\
  }\href {\doibase 10.1103/PhysRevD.80.052002} {\bibfield  {journal} {\bibinfo
  {journal} {Phys.Rev.}\ }\textbf {\bibinfo {volume} {D80}},\ \bibinfo {pages}
  {052002} (\bibinfo {year} {2009})},\ \Eprint {http://arxiv.org/abs/0905.4778}
  {arXiv:0905.4778 [hep-ex]} \BibitemShut {NoStop}%
%%CITATION = ARXIV:0905.4778;%%
\bibitem [{\citenamefont {Uehara}\ \emph {et~al.}(2012)\citenamefont {Uehara}
  \emph {et~al.}}]{Uehara:2012ag}%
  \BibitemOpen
  \bibfield  {author} {\bibinfo {author} {\bibfnamefont {S.}~\bibnamefont
  {Uehara}} \emph {et~al.} (\bibinfo {collaboration} {Belle Collaboration}),\
  }\href@noop {} {\  (\bibinfo {year} {2012})},\ \Eprint
  {http://arxiv.org/abs/1205.3249} {arXiv:1205.3249 [hep-ex]} \BibitemShut
  {NoStop}%
%%CITATION = ARXIV:1205.3249;%%
\bibitem [{\citenamefont {Brodsky}\ and\ \citenamefont
  {Farrar}(1973)}]{Brodsky:1973kr}%
  \BibitemOpen
  \bibfield  {author} {\bibinfo {author} {\bibfnamefont {S.~J.}\ \bibnamefont
  {Brodsky}}\ and\ \bibinfo {author} {\bibfnamefont {G.~R.}\ \bibnamefont
  {Farrar}},\ }\href {\doibase 10.1103/PhysRevLett.31.1153} {\bibfield
  {journal} {\bibinfo  {journal} {Phys. Rev. Lett.}\ }\textbf {\bibinfo
  {volume} {31}},\ \bibinfo {pages} {1153} (\bibinfo {year}
  {1973})}\BibitemShut {NoStop}%
%%CITATION = PRLTA,31,1153;%%
\bibitem [{\citenamefont {Brodsky}\ and\ \citenamefont
  {Farrar}(1975)}]{Brodsky:1974vy}%
  \BibitemOpen
  \bibfield  {author} {\bibinfo {author} {\bibfnamefont {S.~J.}\ \bibnamefont
  {Brodsky}}\ and\ \bibinfo {author} {\bibfnamefont {G.~R.}\ \bibnamefont
  {Farrar}},\ }\href {\doibase 10.1103/PhysRevD.11.1309} {\bibfield  {journal}
  {\bibinfo  {journal} {Phys. Rev.}\ }\textbf {\bibinfo {volume} {D11}},\
  \bibinfo {pages} {1309} (\bibinfo {year} {1975})}\BibitemShut {NoStop}%
%%CITATION = PHRVA,D11,1309;%%
\bibitem [{\citenamefont {Farrar}\ and\ \citenamefont
  {Jackson}(1979)}]{Farrar:1979aw}%
  \BibitemOpen
  \bibfield  {author} {\bibinfo {author} {\bibfnamefont {G.~R.}\ \bibnamefont
  {Farrar}}\ and\ \bibinfo {author} {\bibfnamefont {D.~R.}\ \bibnamefont
  {Jackson}},\ }\href {\doibase 10.1103/PhysRevLett.43.246} {\bibfield
  {journal} {\bibinfo  {journal} {Phys. Rev. Lett.}\ }\textbf {\bibinfo
  {volume} {43}},\ \bibinfo {pages} {246} (\bibinfo {year} {1979})}\BibitemShut
  {NoStop}%
%%CITATION = PRLTA,43,246;%%
\bibitem [{\citenamefont {Radyushkin}(1977)}]{Radyushkin:1977gp}%
  \BibitemOpen
  \bibfield  {author} {\bibinfo {author} {\bibfnamefont {A.~V.}\ \bibnamefont
  {Radyushkin}},\ }\href@noop {} {\  (\bibinfo {year} {1977})},\ \Eprint
  {http://arxiv.org/abs/hep-ph/0410276} {arXiv:hep-ph/0410276} \BibitemShut
  {NoStop}%
%%CITATION = HEP-PH/0410276;%%
\bibitem [{\citenamefont {Efremov}\ and\ \citenamefont
  {Radyushkin}(1980{\natexlab{a}})}]{Efremov:1978rn}%
  \BibitemOpen
  \bibfield  {author} {\bibinfo {author} {\bibfnamefont {A.~V.}\ \bibnamefont
  {Efremov}}\ and\ \bibinfo {author} {\bibfnamefont {A.~V.}\ \bibnamefont
  {Radyushkin}},\ }\href {\doibase 10.1007/BF01032111} {\bibfield  {journal}
  {\bibinfo  {journal} {Theor. Math. Phys.}\ }\textbf {\bibinfo {volume}
  {42}},\ \bibinfo {pages} {97} (\bibinfo {year}
  {1980}{\natexlab{a}})}\BibitemShut {NoStop}%
%%CITATION = TMPHA,42,97;%%
\bibitem [{\citenamefont {Efremov}\ and\ \citenamefont
  {Radyushkin}(1980{\natexlab{b}})}]{Efremov:1979qk}%
  \BibitemOpen
  \bibfield  {author} {\bibinfo {author} {\bibfnamefont {A.~V.}\ \bibnamefont
  {Efremov}}\ and\ \bibinfo {author} {\bibfnamefont {A.~V.}\ \bibnamefont
  {Radyushkin}},\ }\href {\doibase 10.1016/0370-2693(80)90869-2} {\bibfield
  {journal} {\bibinfo  {journal} {Phys. Lett.}\ }\textbf {\bibinfo {volume}
  {B94}},\ \bibinfo {pages} {245} (\bibinfo {year}
  {1980}{\natexlab{b}})}\BibitemShut {NoStop}%
%%CITATION = PHLTA,B94,245;%%
\bibitem [{\citenamefont {Bryman}\ \emph {et~al.}(1982)\citenamefont {Bryman},
  \citenamefont {Depommier},\ and\ \citenamefont {Leroy}}]{Bryman:1982et}%
  \BibitemOpen
  \bibfield  {author} {\bibinfo {author} {\bibfnamefont {D.}~\bibnamefont
  {Bryman}}, \bibinfo {author} {\bibfnamefont {P.}~\bibnamefont {Depommier}}, \
  and\ \bibinfo {author} {\bibfnamefont {C.}~\bibnamefont {Leroy}},\ }\href
  {\doibase 10.1016/0370-1573(82)90162-4} {\bibfield  {journal} {\bibinfo
  {journal} {Phys.Rept.}\ }\textbf {\bibinfo {volume} {88}},\ \bibinfo {pages}
  {151} (\bibinfo {year} {1982})}\BibitemShut {NoStop}%
%%CITATION = PRPLC,88,151;%%
\bibitem [{\citenamefont {et~al. (Particle Data~Group)}(2012)}]{PDG2012}%
  \BibitemOpen
  \bibfield  {author} {\bibinfo {author} {\bibfnamefont {J.~B.}\ \bibnamefont
  {et~al. (Particle Data~Group)}},\ }\href@noop {} {\bibfield  {journal}
  {\bibinfo  {journal} {Phys. Rev. D}\ }\textbf {\bibinfo {volume} {86}}
  (\bibinfo {year} {2012})}\BibitemShut {NoStop}%
\bibitem [{\citenamefont {Knecht}\ and\ \citenamefont
  {Nyffeler}(2002)}]{Knecht:2001qf}%
  \BibitemOpen
  \bibfield  {author} {\bibinfo {author} {\bibfnamefont {M.}~\bibnamefont
  {Knecht}}\ and\ \bibinfo {author} {\bibfnamefont {A.}~\bibnamefont
  {Nyffeler}},\ }\href {\doibase 10.1103/PhysRevD.65.073034} {\bibfield
  {journal} {\bibinfo  {journal} {Phys.Rev.}\ }\textbf {\bibinfo {volume}
  {D65}},\ \bibinfo {pages} {073034} (\bibinfo {year} {2002})},\ \Eprint
  {http://arxiv.org/abs/hep-ph/0111058} {arXiv:hep-ph/0111058 [hep-ph]}
  \BibitemShut {NoStop}%
%%CITATION = HEP-PH/0111058;%%
\bibitem [{\citenamefont {Donoghue}\ and\ \citenamefont
  {Leutwyler}(1991)}]{Donoghue:1991qv}%
  \BibitemOpen
  \bibfield  {author} {\bibinfo {author} {\bibfnamefont {J.~F.}\ \bibnamefont
  {Donoghue}}\ and\ \bibinfo {author} {\bibfnamefont {H.}~\bibnamefont
  {Leutwyler}},\ }\href {\doibase 10.1007/BF01560453} {\bibfield  {journal}
  {\bibinfo  {journal} {Z. Phys.}\ }\textbf {\bibinfo {volume} {C52}},\
  \bibinfo {pages} {343} (\bibinfo {year} {1991})}\BibitemShut {NoStop}%
%%CITATION = ZEPYA,C52,343;%%
\bibitem [{\citenamefont {Brommel}(2007)}]{Brommel:PhD}%
  \BibitemOpen
  \bibfield  {author} {\bibinfo {author} {\bibfnamefont {D.}~\bibnamefont
  {Brommel}},\ }\emph {\bibinfo {title} {{Pion structute from the lattice}}},\
  \href@noop {} {Ph.D. thesis},\ \bibinfo  {school} {{University of
  Regensburg}}, \bibinfo {address} {{Regensburg, Germany}} (\bibinfo {year}
  {2007}),\ \bibinfo {note} {dESY-THESIS-2007-023}\BibitemShut {NoStop}%
\bibitem [{\citenamefont {Brommel}\ \emph {et~al.}(2008)\citenamefont {Brommel}
  \emph {et~al.}}]{Brommel:2007xd}%
  \BibitemOpen
  \bibfield  {author} {\bibinfo {author} {\bibfnamefont {D.}~\bibnamefont
  {Brommel}} \emph {et~al.} (\bibinfo {collaboration} {QCDSF}),\ }\href
  {\doibase 10.1103/PhysRevLett.101.122001} {\bibfield  {journal} {\bibinfo
  {journal} {Phys. Rev. Lett.}\ }\textbf {\bibinfo {volume} {101}},\ \bibinfo
  {pages} {122001} (\bibinfo {year} {2008})},\ \Eprint
  {http://arxiv.org/abs/0708.2249} {arXiv:0708.2249 [hep-lat]} \BibitemShut
  {NoStop}%
%%CITATION = 0708.2249;%%
\bibitem [{\citenamefont {Dubnicka}\ \emph {et~al.}(2003)\citenamefont
  {Dubnicka}, \citenamefont {Dubnickova},\ and\ \citenamefont
  {Weisenpacher}}]{Dubnicka:2002yp}%
  \BibitemOpen
  \bibfield  {author} {\bibinfo {author} {\bibfnamefont {S.}~\bibnamefont
  {Dubnicka}}, \bibinfo {author} {\bibfnamefont {A.-Z.}\ \bibnamefont
  {Dubnickova}}, \ and\ \bibinfo {author} {\bibfnamefont {P.}~\bibnamefont
  {Weisenpacher}},\ }\href {\doibase 10.1088/0954-3899/29/2/316} {\bibfield
  {journal} {\bibinfo  {journal} {J. Phys.}\ }\textbf {\bibinfo {volume}
  {G29}},\ \bibinfo {pages} {405} (\bibinfo {year} {2003})},\ \Eprint
  {http://arxiv.org/abs/hep-ph/0208051} {arXiv:hep-ph/0208051} \BibitemShut
  {NoStop}%
%%CITATION = HEP-PH/0208051;%%
\bibitem [{\citenamefont {Faessler}\ \emph {et~al.}(2010)\citenamefont
  {Faessler}, \citenamefont {Krivoruchenko},\ and\ \citenamefont
  {Martemyanov}}]{Faessler:2009tn}%
  \BibitemOpen
  \bibfield  {author} {\bibinfo {author} {\bibfnamefont {A.}~\bibnamefont
  {Faessler}}, \bibinfo {author} {\bibfnamefont {M.~I.}\ \bibnamefont
  {Krivoruchenko}}, \ and\ \bibinfo {author} {\bibfnamefont {B.~V.}\
  \bibnamefont {Martemyanov}},\ }\href {\doibase 10.1103/PhysRevC.82.038201}
  {\bibfield  {journal} {\bibinfo  {journal} {Phys. Rev.}\ }\textbf {\bibinfo
  {volume} {C82}},\ \bibinfo {pages} {038201} (\bibinfo {year} {2010})},\
  \Eprint {http://arxiv.org/abs/0910.5589} {arXiv:0910.5589 [hep-ph]}
  \BibitemShut {NoStop}%
%%CITATION = 0910.5589;%%
\bibitem [{\citenamefont {Mergell}\ \emph {et~al.}(1996)\citenamefont
  {Mergell}, \citenamefont {Meissner},\ and\ \citenamefont
  {Drechsel}}]{Mergell:1995bf}%
  \BibitemOpen
  \bibfield  {author} {\bibinfo {author} {\bibfnamefont {P.}~\bibnamefont
  {Mergell}}, \bibinfo {author} {\bibfnamefont {U.~G.}\ \bibnamefont
  {Meissner}}, \ and\ \bibinfo {author} {\bibfnamefont {D.}~\bibnamefont
  {Drechsel}},\ }\href {\doibase 10.1016/0375-9474(95)00339-8} {\bibfield
  {journal} {\bibinfo  {journal} {Nucl. Phys.}\ }\textbf {\bibinfo {volume}
  {A596}},\ \bibinfo {pages} {367} (\bibinfo {year} {1996})},\ \Eprint
  {http://arxiv.org/abs/hep-ph/9506375} {arXiv:hep-ph/9506375} \BibitemShut
  {NoStop}%
%%CITATION = HEP-PH/9506375;%%
\bibitem [{\citenamefont {Kelly}(2004)}]{Kelly:2004hm}%
  \BibitemOpen
  \bibfield  {author} {\bibinfo {author} {\bibfnamefont {J.~J.}\ \bibnamefont
  {Kelly}},\ }\href {\doibase 10.1103/PhysRevC.70.068202} {\bibfield  {journal}
  {\bibinfo  {journal} {Phys. Rev.}\ }\textbf {\bibinfo {volume} {C70}},\
  \bibinfo {pages} {068202} (\bibinfo {year} {2004})}\BibitemShut {NoStop}%
%%CITATION = PHRVA,C70,068202;%%
\bibitem [{\citenamefont {Machleidt}\ \emph {et~al.}(1987)\citenamefont
  {Machleidt}, \citenamefont {Holinde},\ and\ \citenamefont
  {Elster}}]{Machleidt:1987hj}%
  \BibitemOpen
  \bibfield  {author} {\bibinfo {author} {\bibfnamefont {R.}~\bibnamefont
  {Machleidt}}, \bibinfo {author} {\bibfnamefont {K.}~\bibnamefont {Holinde}},
  \ and\ \bibinfo {author} {\bibfnamefont {C.}~\bibnamefont {Elster}},\ }\href
  {\doibase 10.1016/S0370-1573(87)80002-9} {\bibfield  {journal} {\bibinfo
  {journal} {Phys.Rept.}\ }\textbf {\bibinfo {volume} {149}},\ \bibinfo {pages}
  {1} (\bibinfo {year} {1987})}\BibitemShut {NoStop}%
%%CITATION = PRPLC,149,1;%%
\bibitem [{\citenamefont {Arrington}\ \emph {et~al.}(2007)\citenamefont
  {Arrington}, \citenamefont {Melnitchouk},\ and\ \citenamefont
  {Tjon}}]{Arrington:2007ux}%
  \BibitemOpen
  \bibfield  {author} {\bibinfo {author} {\bibfnamefont {J.}~\bibnamefont
  {Arrington}}, \bibinfo {author} {\bibfnamefont {W.}~\bibnamefont
  {Melnitchouk}}, \ and\ \bibinfo {author} {\bibfnamefont {J.}~\bibnamefont
  {Tjon}},\ }\href {\doibase 10.1103/PhysRevC.76.035205} {\bibfield  {journal}
  {\bibinfo  {journal} {Phys.Rev.}\ }\textbf {\bibinfo {volume} {C76}},\
  \bibinfo {pages} {035205} (\bibinfo {year} {2007})},\ \Eprint
  {http://arxiv.org/abs/0707.1861} {arXiv:0707.1861 [nucl-ex]} \BibitemShut
  {NoStop}%
%%CITATION = ARXIV:0707.1861;%%
\bibitem [{\citenamefont {Perdrisat}\ \emph {et~al.}(2007)\citenamefont
  {Perdrisat}, \citenamefont {Punjabi},\ and\ \citenamefont
  {Vanderhaeghen}}]{Perdrisat:2006hj}%
  \BibitemOpen
  \bibfield  {author} {\bibinfo {author} {\bibfnamefont {C.}~\bibnamefont
  {Perdrisat}}, \bibinfo {author} {\bibfnamefont {V.}~\bibnamefont {Punjabi}},
  \ and\ \bibinfo {author} {\bibfnamefont {M.}~\bibnamefont {Vanderhaeghen}},\
  }\href@noop {} {\bibfield  {journal} {\bibinfo  {journal}
  {Prog.Part.Nucl.Phys.}\ }\textbf {\bibinfo {volume} {59}},\ \bibinfo {pages}
  {694} (\bibinfo {year} {2007})}\BibitemShut {NoStop}%
\bibitem [{\citenamefont {Alexandrou}\ \emph {et~al.}(2011)\citenamefont
  {Alexandrou} \emph {et~al.}}]{Alexandrou:2010hf}%
  \BibitemOpen
  \bibfield  {author} {\bibinfo {author} {\bibfnamefont {C.}~\bibnamefont
  {Alexandrou}} \emph {et~al.} (\bibinfo {collaboration} {ETM Collaboration}),\
  }\href {\doibase 10.1103/PhysRevD.83.045010} {\bibfield  {journal} {\bibinfo
  {journal} {Phys.Rev.}\ }\textbf {\bibinfo {volume} {D83}},\ \bibinfo {pages}
  {045010} (\bibinfo {year} {2011})},\ \Eprint {http://arxiv.org/abs/1012.0857}
  {arXiv:1012.0857 [hep-lat]} \BibitemShut {NoStop}%
%%CITATION = ARXIV:1012.0857;%%
\bibitem [{\citenamefont {Bernard}\ \emph {et~al.}(2002)\citenamefont
  {Bernard}, \citenamefont {Elouadrhiri},\ and\ \citenamefont
  {Meissner}}]{Bernard:2001rs}%
  \BibitemOpen
  \bibfield  {author} {\bibinfo {author} {\bibfnamefont {V.}~\bibnamefont
  {Bernard}}, \bibinfo {author} {\bibfnamefont {L.}~\bibnamefont
  {Elouadrhiri}}, \ and\ \bibinfo {author} {\bibfnamefont {U.}~\bibnamefont
  {Meissner}},\ }\href {\doibase 10.1088/0954-3899/28/1/201} {\bibfield
  {journal} {\bibinfo  {journal} {J.Phys.G}\ }\textbf {\bibinfo {volume}
  {G28}},\ \bibinfo {pages} {R1} (\bibinfo {year} {2002})},\ \Eprint
  {http://arxiv.org/abs/hep-ph/0107088} {arXiv:hep-ph/0107088 [hep-ph]}
  \BibitemShut {NoStop}%
\bibitem [{\citenamefont {Dominguez}(1977)}]{Dominguez:1976ut}%
  \BibitemOpen
  \bibfield  {author} {\bibinfo {author} {\bibfnamefont {C.}~\bibnamefont
  {Dominguez}},\ }\href {\doibase 10.1103/PhysRevD.15.1350} {\bibfield
  {journal} {\bibinfo  {journal} {Phys.Rev.}\ }\textbf {\bibinfo {volume}
  {D15}},\ \bibinfo {pages} {1350} (\bibinfo {year} {1977})}\BibitemShut
  {NoStop}%
\bibitem [{\citenamefont {Alvegard}\ and\ \citenamefont
  {Kogerler}(1979)}]{Alvegard:1979ui}%
  \BibitemOpen
  \bibfield  {author} {\bibinfo {author} {\bibfnamefont {C.}~\bibnamefont
  {Alvegard}}\ and\ \bibinfo {author} {\bibfnamefont {R.}~\bibnamefont
  {Kogerler}},\ }\href {\doibase 10.1007/BF01474131} {\bibfield  {journal}
  {\bibinfo  {journal} {Z.Phys.}\ }\textbf {\bibinfo {volume} {C2}},\ \bibinfo
  {pages} {173} (\bibinfo {year} {1979})}\BibitemShut {NoStop}%
\bibitem [{\citenamefont {Brodsky}\ \emph {et~al.}(1981)\citenamefont
  {Brodsky}, \citenamefont {Lepage},\ and\ \citenamefont
  {Zaidi}}]{Brodsky:1980sx}%
  \BibitemOpen
  \bibfield  {author} {\bibinfo {author} {\bibfnamefont {S.~J.}\ \bibnamefont
  {Brodsky}}, \bibinfo {author} {\bibfnamefont {G.}~\bibnamefont {Lepage}}, \
  and\ \bibinfo {author} {\bibfnamefont {S.}~\bibnamefont {Zaidi}},\ }\href
  {\doibase 10.1103/PhysRevD.23.1152} {\bibfield  {journal} {\bibinfo
  {journal} {Phys.Rev.}\ }\textbf {\bibinfo {volume} {D23}},\ \bibinfo {pages}
  {1152} (\bibinfo {year} {1981})}\BibitemShut {NoStop}%
\bibitem [{\citenamefont {Carlson}\ and\ \citenamefont
  {Poor}(1986)}]{Carlson:1985zu}%
  \BibitemOpen
  \bibfield  {author} {\bibinfo {author} {\bibfnamefont {C.~E.}\ \bibnamefont
  {Carlson}}\ and\ \bibinfo {author} {\bibfnamefont {J.}~\bibnamefont {Poor}},\
  }\href {\doibase 10.1103/PhysRevD.34.1478} {\bibfield  {journal} {\bibinfo
  {journal} {Phys.Rev.}\ }\textbf {\bibinfo {volume} {D34}},\ \bibinfo {pages}
  {1478} (\bibinfo {year} {1986})}\BibitemShut {NoStop}%
\bibitem [{\citenamefont {Carlson}\ and\ \citenamefont
  {Poor}(1987)}]{Carlson:1987en}%
  \BibitemOpen
  \bibfield  {author} {\bibinfo {author} {\bibfnamefont {C.~E.}\ \bibnamefont
  {Carlson}}\ and\ \bibinfo {author} {\bibfnamefont {J.}~\bibnamefont {Poor}},\
  }\href {\doibase 10.1103/PhysRevD.36.2169} {\bibfield  {journal} {\bibinfo
  {journal} {Phys.Rev.}\ }\textbf {\bibinfo {volume} {D36}},\ \bibinfo {pages}
  {2169} (\bibinfo {year} {1987})}\BibitemShut {NoStop}%
\bibitem [{\citenamefont {Gockeler}\ \emph {et~al.}(2004)\citenamefont
  {Gockeler} \emph {et~al.}}]{Gockeler:2003jfa}%
  \BibitemOpen
  \bibfield  {author} {\bibinfo {author} {\bibfnamefont {M.}~\bibnamefont
  {Gockeler}} \emph {et~al.} (\bibinfo {collaboration} {QCDSF}),\ }\href
  {\doibase 10.1103/PhysRevLett.92.042002} {\bibfield  {journal} {\bibinfo
  {journal} {Phys. Rev. Lett.}\ }\textbf {\bibinfo {volume} {92}},\ \bibinfo
  {pages} {042002} (\bibinfo {year} {2004})},\ \Eprint
  {http://arxiv.org/abs/hep-ph/0304249} {arXiv:hep-ph/0304249} \BibitemShut
  {NoStop}%
%%CITATION = HEP-PH/0304249;%%
\bibitem [{\citenamefont {Hagler}(2007)}]{Hagler:2007hu}%
  \BibitemOpen
  \bibfield  {author} {\bibinfo {author} {\bibfnamefont {P.}~\bibnamefont
  {Hagler}},\ }\href@noop {} {\  (\bibinfo {year} {2007})},\ \Eprint
  {http://arxiv.org/abs/arXiv:0711.0819 [hep-lat]} {arXiv:0711.0819 [hep-lat]}
  \BibitemShut {NoStop}%
%%CITATION = ARXIV:0711.0819;%%
\bibitem [{\citenamefont {Anisovich}\ \emph {et~al.}(2012)\citenamefont
  {Anisovich}, \citenamefont {Bugg}, \citenamefont {Nikonov}, \citenamefont
  {Sarantsev},\ and\ \citenamefont {Sarantsev}}]{Anisovich:2011in}%
  \BibitemOpen
  \bibfield  {author} {\bibinfo {author} {\bibfnamefont {A.}~\bibnamefont
  {Anisovich}}, \bibinfo {author} {\bibfnamefont {D.}~\bibnamefont {Bugg}},
  \bibinfo {author} {\bibfnamefont {V.}~\bibnamefont {Nikonov}}, \bibinfo
  {author} {\bibfnamefont {A.}~\bibnamefont {Sarantsev}}, \ and\ \bibinfo
  {author} {\bibfnamefont {V.}~\bibnamefont {Sarantsev}},\ }\href {\doibase
  10.1103/PhysRevD.85.014001} {\bibfield  {journal} {\bibinfo  {journal}
  {Phys.Rev.}\ }\textbf {\bibinfo {volume} {D85}},\ \bibinfo {pages} {014001}
  (\bibinfo {year} {2012})},\ \Eprint {http://arxiv.org/abs/1110.4333}
  {arXiv:1110.4333 [hep-ex]} \BibitemShut {NoStop}%
%%CITATION = ARXIV:1110.4333;%%
\bibitem [{\citenamefont {Nieves}\ \emph {et~al.}(2011)\citenamefont {Nieves},
  \citenamefont {Pich},\ and\ \citenamefont {Ruiz~Arriola}}]{Nieves:2011gb}%
  \BibitemOpen
  \bibfield  {author} {\bibinfo {author} {\bibfnamefont {J.}~\bibnamefont
  {Nieves}}, \bibinfo {author} {\bibfnamefont {A.}~\bibnamefont {Pich}}, \ and\
  \bibinfo {author} {\bibfnamefont {E.}~\bibnamefont {Ruiz~Arriola}},\ }\href
  {\doibase 10.1103/PhysRevD.84.096002} {\bibfield  {journal} {\bibinfo
  {journal} {Phys.Rev.}\ }\textbf {\bibinfo {volume} {D84}},\ \bibinfo {pages}
  {096002} (\bibinfo {year} {2011})},\ \Eprint {http://arxiv.org/abs/1107.3247}
  {arXiv:1107.3247 [hep-ph]} \BibitemShut {NoStop}%
%%CITATION = ARXIV:1107.3247;%%
\bibitem [{\citenamefont {Leinweber}\ and\ \citenamefont
  {Cohen}(1994)}]{Leinweber:1993yw}%
  \BibitemOpen
  \bibfield  {author} {\bibinfo {author} {\bibfnamefont {D.~B.}\ \bibnamefont
  {Leinweber}}\ and\ \bibinfo {author} {\bibfnamefont {T.~D.}\ \bibnamefont
  {Cohen}},\ }\href {\doibase 10.1103/PhysRevD.49.3512} {\bibfield  {journal}
  {\bibinfo  {journal} {Phys.Rev.}\ }\textbf {\bibinfo {volume} {D49}},\
  \bibinfo {pages} {3512} (\bibinfo {year} {1994})},\ \Eprint
  {http://arxiv.org/abs/hep-ph/9307261} {arXiv:hep-ph/9307261 [hep-ph]}
  \BibitemShut {NoStop}%
%%CITATION = HEP-PH/9307261;%%
\end{thebibliography}
\end{document}